\documentclass[10pt,journal,compsoc]{IEEEtran}

\ifCLASSOPTIONcompsoc
  \usepackage[nocompress]{cite}
\else
  \usepackage{cite}
\fi

\usepackage{cite}
\usepackage{amsmath,amssymb,amsfonts}
\usepackage{algorithmic}
\usepackage{graphicx}
\usepackage{color,array}
\usepackage{cite}
\usepackage{amsmath,amssymb,amsfonts}
\usepackage{algorithmic}
\usepackage{textcomp}
\usepackage{braket}
\usepackage{bm}
\usepackage{times}
\usepackage{epsfig}
\usepackage{graphicx}
\usepackage{amsmath}
\usepackage{multirow}
\usepackage{amssymb}
\usepackage{subfig}
\usepackage{graphicx}
\usepackage{float}
\usepackage{booktabs}
\usepackage{textcomp}
\usepackage{adjustbox}\usepackage{mathtools}
\usepackage{array}
\usepackage{makecell,booktabs}
\usepackage{multirow}
\usepackage{stackengine}
\usepackage{algorithm}
\usepackage{graphicx}
\usepackage{textcomp}
\usepackage{wrapfig,lipsum,booktabs}
\usepackage{soul}

\ifCLASSINFOpdf
\else
\fi

\begin{document}
%
\title{Quantum Embedding Search for Quantum Machine Learning}
%
%
%
%

\author{Nam Nguyen,~\IEEEmembership{Student Member,~IEEE,}
        and Kwang-Chen Chen,~\IEEEmembership{Fellow,~IEEE}
\IEEEcompsocitemizethanks{\IEEEcompsocthanksitem M. The authors are with the Department of Electrical Engineering, University of South Florida, Tampa, FL 33620.\protect\\
E-mail: \{namnguyen2,kwangcheng\}@usf.edu}
\thanks{}}

%
%

\markboth{Quantum Embedding Search for Quantum Machine Learning}%
{Nguyen \MakeLowercase{\textit{et al.}}: Quantum Embedding Search for Quantum Machine Learning}

\IEEEtitleabstractindextext{%
\begin{abstract}
This paper introduces an automated search algorithm (QES, pronounced as "quest"), \color{black} which derives optimal design of entangling layout for supervised quantum machine learning. First, we establish the connection between the structures of entanglement using CNOT gates and the representations of directed multi-graphs, enabling a well-defined search space. The proposed encoding scheme of quantum entanglement as genotype vectors bridges the ansatz optimization and classical machine learning, allowing efficient search on any well-defined search space\color{black}. Second, we instigate the entanglement level to reduce the cardinality of the search space to a feasible size for practical implementations. Finally, we mitigate the cost of evaluating the true loss function by using surrogate models via sequential model-based optimization. We demonstrate the feasibility of our proposed approach on simulated and \color{black} bench-marking datasets, including Iris, Wine and Breast Cancer datasets\color{black}, which empirically shows that found quantum embedding architecture by QES outperforms manual designs in term of the predictive performance. 
\end{abstract}

\begin{IEEEkeywords}
Ansatz Optimization, Quantum Embeddings, Quantum Machine Learning, Quantum Logic Gates.
\end{IEEEkeywords}}

\maketitle

\IEEEdisplaynontitleabstractindextext

%
\IEEEpeerreviewmaketitle

\ifCLASSOPTIONcompsoc
\IEEEraisesectionheading{\section{Introduction}\label{sec:introduction}}
\begin{figure*}[t]
    \centering
    \includegraphics[width = 0.7\textwidth]{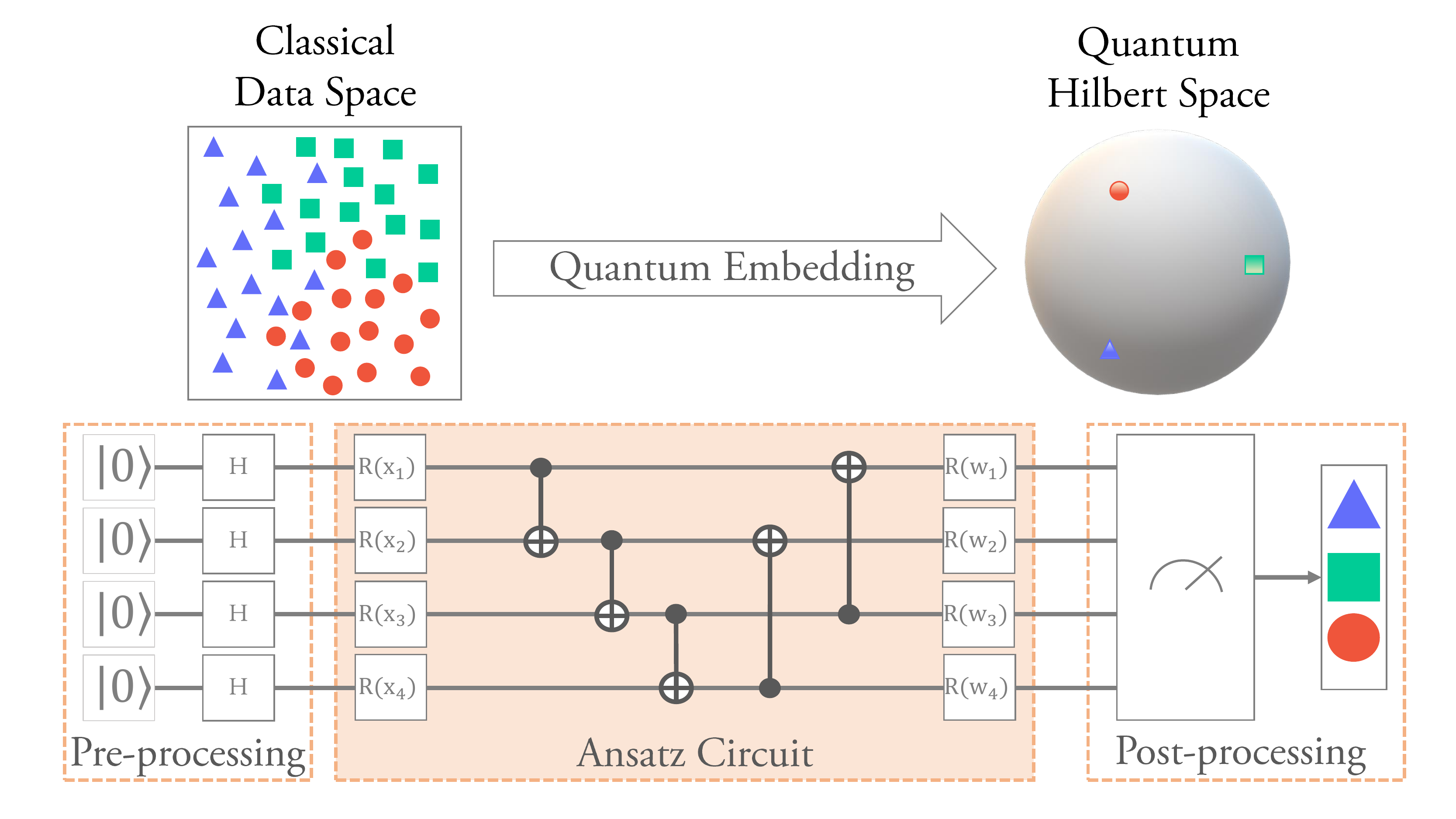}
    \caption{(Top) \textbf{Quantum embeddings for supervised quantum machine learning}. Similar to kernel methods, quantum embedding transforms observations in classical data space into quantum Hilbert space of quantum states, which the inner product of quantum representations can represent. (Bottom) \textbf{Architecture of a quantum machine learning model formed by selected set of quantum circuits}. The ansatz circuit plays an crucial role in the circuit model, enabling learning model's weights $\bm{w}$ accordingly to input data $\bm{x}$.}
    \label{fig:qe-qml}
\end{figure*}
Quantum machine learning is view as a potential advancement of quantum computing in the Noisy Intermediate-Scale Quantum (NISQ) era. As the validity of near-term quantum devices, quantum machine learning poses exciting advantages over classical counterparts. The potential quantum advantage can be addressed based on the geometric test over the input data space, followed by the complexity test for specific functions\cite{huang2021power}. Although quantum machine learning models are more often than not referred to as quantum neural networks, the terminology might be misleading to some extend. The classical neural networks can transform the original data space into higher or lower dimensional space based on the design of neural architectures. For example, state-of-the-art neural architectures tend to transform high-dimensional inputs such as images into lower-dimensional representations of latent vectors. In contrast, quantum neural networks possess a similar mathematical structure to kernel methods, where input data is embedded into high-dimensional quantum Hilbert space\cite{schuld2021supervised,havlivcek2019supervised,schuld2019quantum}. The quantum representations of input data are the outcome of quantum embedding, which plays a crucial role in the performance of quantum classifiers \cite{lloyd2020quantum}. Such quantum embeddings are quantum model functions, referred to as parameterized quantum circuits\cite{benedetti2019parameterized}, quantum neural networks\cite{farhi2018classification,mcclean2018barren}, or variational circuits\cite{romero2021variational,mcclean2016theory}. The quantum embeddings are often manually designed for specific use-cases, which requires extensive expert knowledge and computational resource. \color{black} From the perspective of deep learning, the classical embeddings aim to transform the inputs into deep representations in the latent space, which commonly has lower dimensionality. For example, convolutional neural networks (CNNs) embeds the input images (considered high-dimensional input) into deep brief features (lower-dimensional representations in the laten space), which enables performing machine learning tasks such as classification or object detection. On the other hand, variational quantum embeddings also transforms the inputs into feature maps; however, the latent space in quantum embeddings is high-dimensional Hilbert space. \cite{lloyd2020quantum} shows that decision boundaries established in the Hilbert space is associated with complex decision boundaries in the input space.

\color{black}
This paper introduces an automated search algorithm, \color{black} which derives optimal design of entangling layout for supervised quantum machine learning. \color{black} First, the proposed work directly addresses the ansatz optimization for emerging quantum machine learning via searching the optimal entanglement layout for ansatz architectures. The novel encoding scheme of entanglement as genotype vectors allows us to leverage ML-based search algorithms for the problem of ansatz optimization, which results in well-performed quantum neural networks\color{black}. Second, we instigate the entanglement level to reduce the cardinality of the search space to a feasible size for practical implementations. Finally, we mitigate the cost of evaluating the true loss function by using surrogate models via sequential model-based optimization. We demonstrate the feasibility of our proposed approach on simulated and \color{black} bench-marking datasets, including Iris, Wine and Breast Cancer datasets\color{black}, which empirically shows that found quantum embedding architecture by QES outperforms manual designs of entanglement in term of the predictive performance. 
\begin{enumerate}
    \item We instigate an efficient encoding scheme of quantum embedding's architectures as directed multi-graphs, which enable us to well-define the search space of the quantum embedding search problem. Moreover, we introduce the constraints over the search space by quantum entanglement level, which reduces the cardinality of the search space to a reasonable size for practical implementations. \color{black}The formulation of quantum entanglement as genotype vectors allows classical ML algorithms efficiently address the ansatz optimization problems.\color{black}
    \item Leveraging the sequential model-based optimization via Tree Parzen Estimator \cite{bergstra2011algorithms} (SMBO-TPE) our search strategy enjoy two-fold advantage: (1) usage of surrogate models enables the approximation of the actual loss function, which significantly reduces computational cost in the optimization, and (2) non-parametric densities in TPE allows us to draw multiple architecture candidates for evaluating the expected improvement of surrogates, which is more computationally effective.
    \item Discovered quantum embedding architectures by QES outperform manual designs in both synthesis and Iris dataset\cite{fisher1936use} while achieving compatible results compared to classical machine learning models.
\end{enumerate}
We organize the paper as: Section~\ref{related-work} summarizes related work in the literature, Section~\ref{qes} discusses our proposed QES algorithm in-depth, Section~\ref{experiments} reports experimental results. Finally, we discuss the implication and threats to the validity of QES in Section~\ref{discusion}.

\section{Related Works}\label{related-work}
\subsection{Automated Architecture Search}

Automated architecture search has drawn significant attention from the ML/DL-related research community. Its motivation is practical but straightforward; that is, there is no universal design of network for all datasets. The main objective of such an algorithm is to find an optimal design for the model's architecture based on pre-defined selection criteria. Initialization of automated architecture search algorithm starts with defining the configuration of the search space. The basic search space structure is known as flat search space, referred to as hyper-parameters optimization. For example, the flat search space of neural architecture search is to find the depth (number of layers), the width (number of initial channels), and the size of kernel. A more complicated formation of search space is cell-based neural architectures, where each neural candidate can be encoded as a directed acyclic graph\cite{zoph2016neural,zoph2018learning}. The search space of our proposed QES is motivated by the latter configuration, which will be discussed hereafter in Section~\ref{qes}. 
Many frameworks have been proposed to tackle the automated architecture search problems. An early solution involves random search\cite{bergstra2012random}, which is often used as the baseline for comparison. The next progression is the development of sequential model-based optimizations, which mitigate the expensive cost of actual loss function by using surrogate models\cite{bergstra2011algorithms,snoek2012practical}. More advanced search strategies have been proposed to tackle the problem, involving reinforcement learning\cite{zoph2016neural}, evolutionary\cite{real2019regularized}, gradient-based with continuous relaxation and bilevel optimization\cite{liu2018darts}, heuristic search with performance prediction\cite{liu2018progressive} and SMBO-TPE\cite{nguyen2021contrastive}.
The biggest challenge in automated architecture search is the computational cost for the search phase. Search strategies such as reinforcement learning and evolutionary take up to $2000-3150$ GPU days to find optimal architecture for the CIFAR-10 dataset\cite{zoph2016neural,zoph2018learning}. Although progression helps to shorten the time complexity of the search procedure to reasonable time and hardware, the expensive computation is inherited from the costly evaluation of the cost function. The same issue appears while training quantum machine learning models in near-term quantum computers and quantum simulators. Recent work solves quantum circuit optimization problems by reinforcement learning with circuit transformation\cite{fosel2021quantum}, which achieves remarkable results. However, such an algorithm still relies on evaluating the true loss function for maximizing the cumulative reward. We find that sequential model-based optimization is another potential solution for quantum embedding search/quantum circuit optimization since the approximation of true loss function by surrogates reduces the computational expense of searching for the optimal quantum embedding architecture.
\subsection{Quantum Machine Learning}
Quantum machine learning has become an emerging technology of quantum computing due to its potential for near-term intermediate-scale quantum hardware. Current literature has witnessed the advantages of quantum machine learning over its classical counterparts given various learning tasks\cite{liu2018quantum,gao2018quantum,dunjko2016quantum,von2018quantum,lloyd2016quantum,lloyd2014quantum}. The primary approach for quantum machine learning is circuit-based models, referred to as variational quantum classifiers\cite{romero2021variational,mcclean2016theory,farhi2014quantum}. Different strategies of classifier in the quantum Hilbert space have been proposed, including linear classifier\cite{mari2020transfer}, bitstrings parity-binary mapping\cite{cappelletti2020polyadic}, Helstrom, and fidelity classifiers\cite{lloyd2020quantum}. Moreover, a strong connection between quantum machine learning and kernel methods has been established in \cite{schuld2021supervised,schuld2019quantum,havlivcek2019supervised}.
The core component of circuit-based quantum machine learning models is the variational (parameterized) circuits called ansatz (plural ansaetze). 
The construction of an ansatz is formed by stacking multiple identical sub-layers, similar to the construction of cell-based neural architecture designs\cite{zoph2016neural}. Although many variational ansaetze have been proposed in the literature\cite{schuld2020circuit,killoran2019continuous,huggins2019towards,farhi2014quantum,farhi2018classification,schuld2019quantum,tacchino2020variational}, there is no general framework to design optimal ansatz for data-specific scenarios or specific use-cases. It is the main motivation for our QES algorithm, which directly tackles the problem of discovering optimal quantum embedding architectures for given datasets of interest.

\color{black}
\subsection{Ansatz Optimization}
Optimizing ansatz circuit plays an indispensable role in designing quantum algorithms for specific tasks in practice. The ansatz optimization problem can be categorized into two main types. In the first category, we perform \textit{circuit simplification} to reduce the computation for quantum hardware. In other words, the local or global structure of ansatz are optimized by being replaced with equivalent but more computationally efficient architectures\cite{nam2018automated,majumdar2021optimizing,fosel2021quantum}. On the other hand, the second categorize of ansatz optimization aim to find the optimal ansatz that yields the best performance on given tasks. In other words, the heuristic search enables well-performed ansatz on specific tasks instead of reduce computation. Our proposed work is in the second category, which aims to find the optimal ansatz for quantum machine learning problems. There are several studies having the same objective with our proposed work, including:\cite{verdon2019learning} introduces the usage of classical neural networks as surrogates to approximate the optimal parameters for tasks such as the Quantum Approximate Optimization Algorithm (QAOA) for MaxCut and Sherrington-Kirkpatrick Ising model or VQE for Hubbard model. \cite{mitarai2018quantum} leverages student-teacher learning (or knowledge distillation) for tuning the circuit parameters such that the ansatz output in pre-chosen output. \cite{ostaszewski2021structure} optimizes VQE for discovering the ground stats of Lithium Hydride and Heisenberg model.
\color{black}

\section{Methodology}\label{qes}
\color{black}
Our proposed QES aims the find the optimal quantum embedding for supervised quantum machine learning under classification tasks. We follow the common design of ansatz used for supervised ML\cite{schuld2021supervised}, which is discussed in Section~\ref{ansatz_qe}. In this formalism, the quantum embedding is simplified into sub-components, including feature-dependent block, entanglement structure between qubits and parameterized rotations (considered as model weights). This layered structure is referred to quantum neural networks (QNNs), which enables feature maps in Hilbert space. Although only consider very restricted design in compare to universal design of ansatz, these designs show promising results towards applications of quantum machine learning in NISQ era. Besides, the optimization of such embeddings remain challenging even in the restricted designs since all combinations of the choices and number for rotation or CNOT gates form a massive search space. Hence, QES reduces the search space by a assumption that the rotation gates are the static components of the ansatz. In other words, our proposed scheme develop the optimal entanglement layout for quantum embeddings. We theoretically show that even in very restrict manner, the search space of finding optimal entanglement layout is massive and exponentially expanded when the number of qubits increases in Section~\ref{search-space}. Thus, the SMBO-TPE for search strategy of ansatz optimization offers several advantage over other competitors such as genetic algorithms or reinforcement learning. First, SMBO leverages surrogates to approximate the true value of fitness function, which significantly reduce the cost for training QNNs in the current implementation. Second, the chosen search strategy enables efficient searching by leveraging the prior knowledge. Finally, with sorted queries in the history, SMBO using TPE can be efficiently scalable when the search space expanded (Section~\ref{smbo}).
\color{black}

\subsection{Quantum Embeddings}\label{ansatz_qe}
Let $\bm{x}\in \mathcal{X}$ be the feature vector in classical data space, quantum embedding is similar to the kernel method since the input feature space is mapped to a high-dimensional Hilbert space \cite{schuld2021supervised}. Mathematically speaking, the mapping is given by
\begin{equation}
    \begin{split}
        \mathcal{X} &\rightarrow \mathcal{H}\\
        \bm{x} &\rightarrow \ket{\bm{\phi(x)}},
    \end{split}
\end{equation}
where $\bm{\phi(x)}$ is the \textit{quantum representations} of original input data and $\mathcal{H}$ is the quantum Hilbert space. In particular, a system of $n$ qubits is corresponding to a vector space $\mathbb{C}^{2^n}$. The measurements on these quantum states yield embedded outputs, that is the representation of the observable in the latent space. In general, the intermediate representations of a quantum state can be written as
\begin{equation}
    g(\bm{x}) = \bra{\bm{x}}\mathcal{Z} \ket{\bm{x}},
\end{equation}
where $\mathcal{Z}$ is the measurement associated with the observable. Our choice for $\mathcal{Z}$ is the expectation values in $\mathbb{Z}$ basis over all qubits, which is $\mathcal{Z} = \prod_{1}^{\otimes n} \sigma_{z}$. State-of-the-art quantum machine learning model\cite{lloyd2020quantum,schuld2021supervised} transforms the expectation in continuous domain to categorical labels by thresholding the outcome. In contrast, we leverage the continuous latent representation of the measurements. The decision boundary of the quantum machine learning model in Figure~\ref{fig:qe-qml} is created by single-layer linear classifier. We will show the power of representation learning from quantum embeddings over classical counterparts with five-time complexity hereafter in Section~\ref{experiments}.

 The layered gat architectures of a quantum embedding includes a stack of multiple circuit ansatz (Figure~\ref{fig:qe-qml}) which results in intractable latent representations for universal quantum computing \cite{lloyd2018quantum}. The Quantum Approximate Optimization Algorithm (QAOA)\cite{farhi2014quantum} inspires the embedding circuit ansatz, which transforms the classical input data into quantum representations. Figure~\ref{fig:qe-qml} shows our assumption for the design of a partially parameterized circuit ansatz, which includes an input-depend unitary block, fixed unitary block and learnable unitary block. Primitive gates of unitary includes Control-rotation gates $R_{\sigma_{1} \in \{X,Y,Z\} }(\bm{x})$, which are parameterized by the input data $\bm{x}$. Then immediate quantum representations $U_{1}\bm{\ket{\phi(x)}}$ are fed forward into the fixed unitary block including multiple entangling patterns active on certain number of qubits. The primitive gates for entanglement establishment is CNOT gate, which offers a highly entangled state over all qubits in the system. Finally, the last block of an ansatz circuit includes control rotation gates $R_{\sigma_{2} \in \{X,Y,Z\} }(\bm{w})$ parameterized by learnable weights $w$. Mathematically, the ansatz acts on $n$-qubits system as
 \begin{equation}
     \ket{x} \rightarrow \Psi(\bm{x},\bm{w})\ket{0}^{\otimes n},
 \end{equation}
 where $\Psi(\bm{x},\bm{w})$ is parameterized unitary transformation of $\bm{w}$ with realizations $\bm{x}$. By backpropagation, the set of weights $\bm{w}$ will be learned to minimize the cost function throughout the training process. Figure~\ref{fig:graph} depicts several manual designs for strongly entangled patterns of an quantum embeddings, which will be considered as baseline comparison for Section~\ref{experiments}.

\textbf{Observations:}

Before further exploration, we would like to address several findings from our observations over the preliminary experiments (Figure~\ref{fig:prelim-results}). 
\begin{enumerate}
    \item Different entangling structures result in varying loss values in the validation set.
    \item Entangling structure is permutation variant. In other words, the order of CNOT gates over qubits significantly impacts the overall performance.
    \item Larger number of CNOT gates does not guarantee higher predictive power of the resulting architecture.
    \item Repetitions of similar entangling connections are possible.
\end{enumerate}
The detail of preliminary experiments on the Iris dataset is given in Appendix~\ref{apd:prelim}, where several statistical tests are conducted to provide statistical evidence for our observations.

\subsection{Search Space Configuration}\label{search-space}
\begin{figure}[t]
    \centering
    \includegraphics[width = 0.48\textwidth]{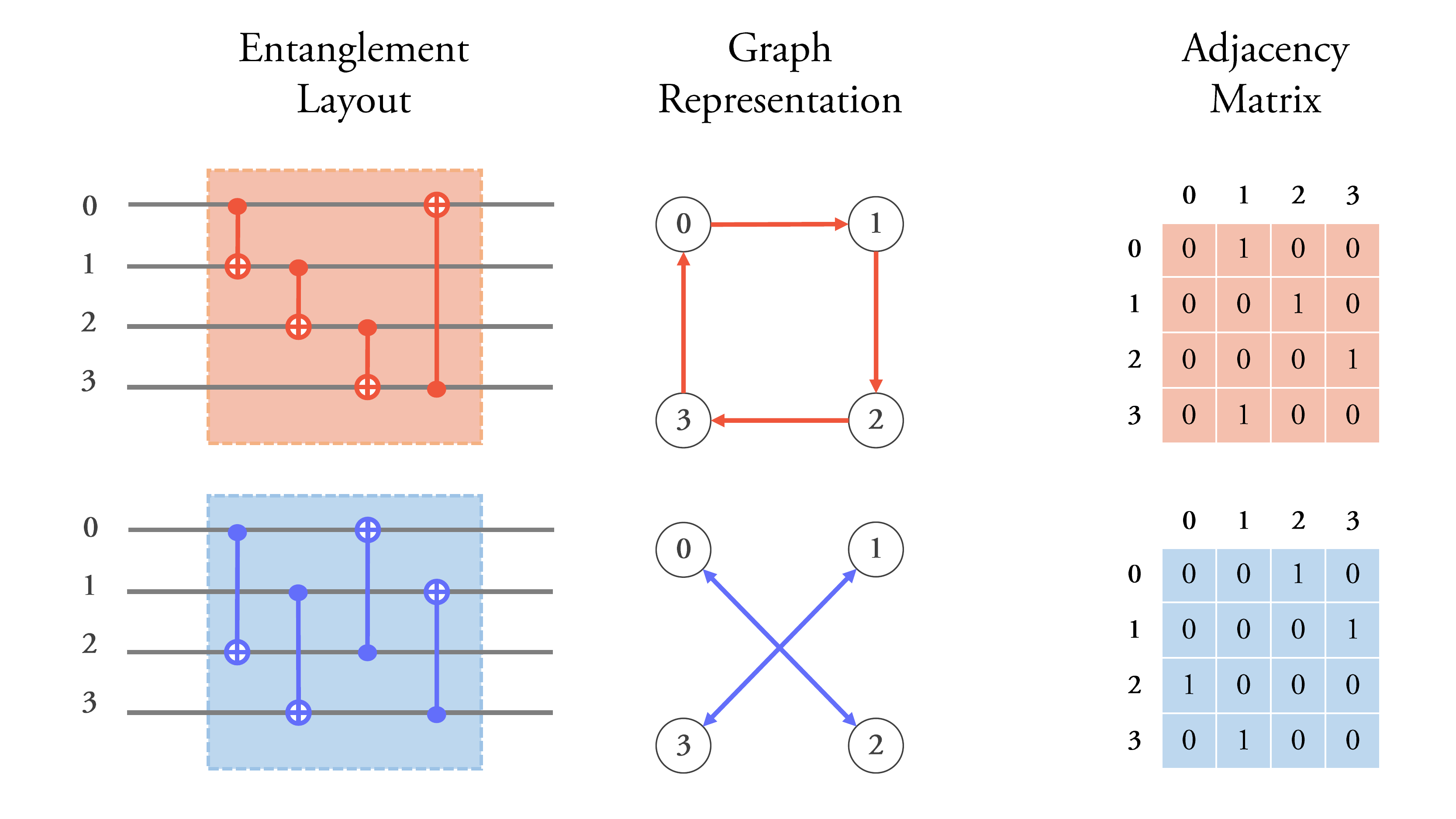}
    \caption{\textbf{Directed multi-graph representation for given entangling structures}. The representation of corresponding graphs are also represented as adjacency matrices, where diagonal entries are $0$ and off-diagonal elements takes value $0$ (absence of CNOT gate) or $1$ (presence of CNOT gate). Moreover, these hand-crated designs are referred as baseline 1 and baseline 2, respectively.}
    \label{fig:graph}
\end{figure}

\begin{figure}[t]
    \centering
    \includegraphics[width = 0.40\textwidth]{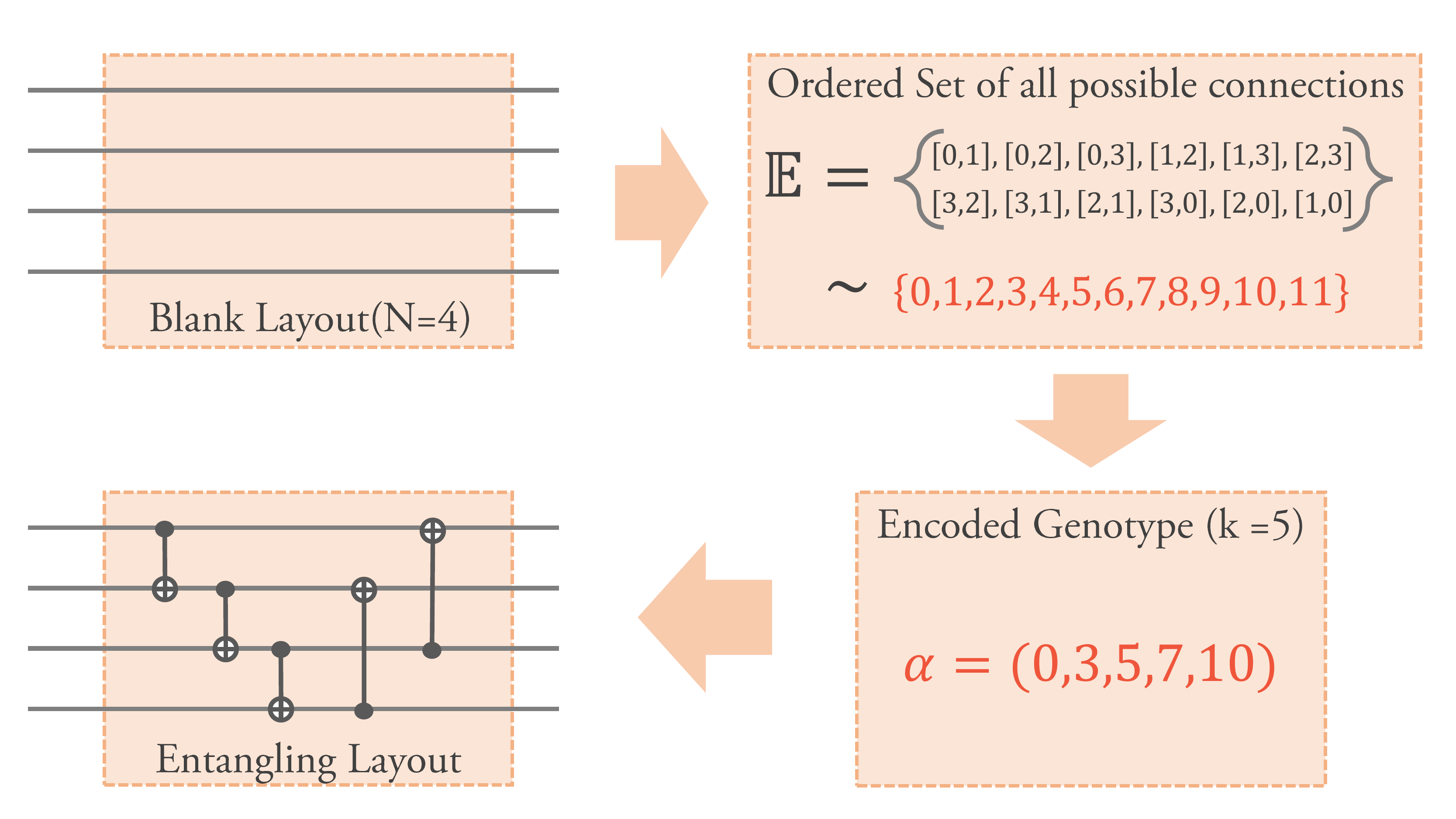}
    \caption{\textbf{Encoding scheme for the search generator}. First, the set of all possible edges is generated considering order of elements. Then "genotype" vector of length $k$ is drawn by the quantum search engine, in which elements are index of corresponding edges.}
    \label{fig:encoding}
\end{figure}

\begin{figure}[t]
    \centering
    \includegraphics[width = 0.40\textwidth]{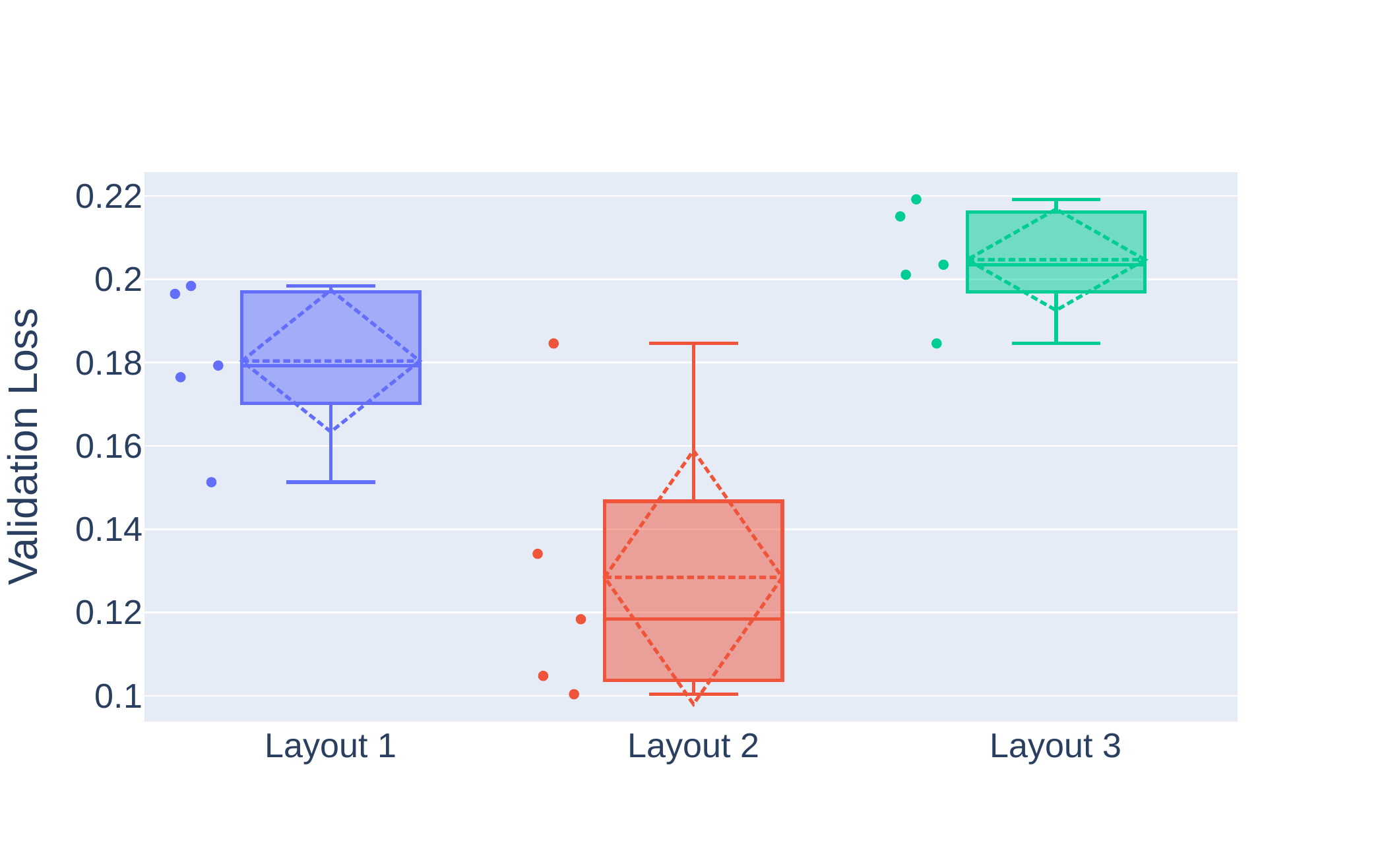}
    \includegraphics[clip, trim = 0cm 6cm 1cm 6cm,width = 0.40\textwidth]{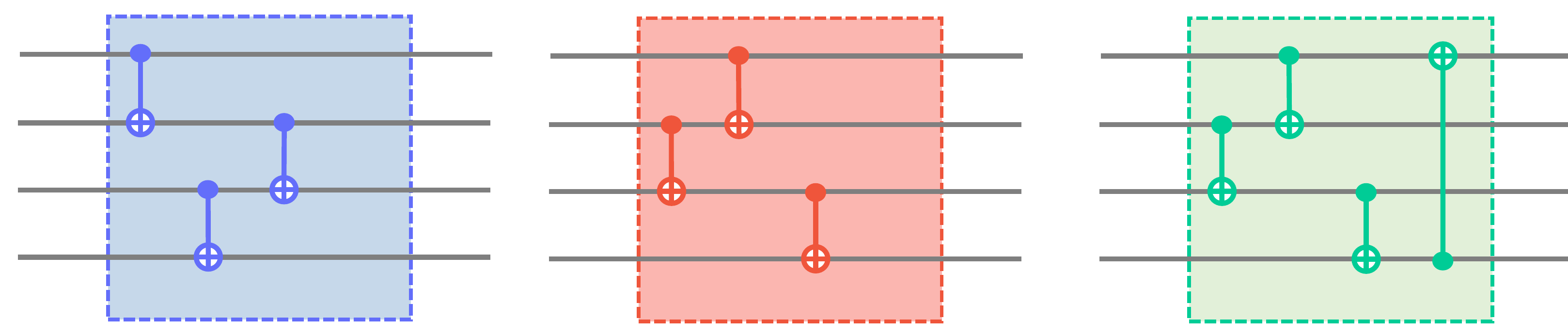}
    \caption{\textbf{Preliminary results of quantum embeddings with different entangling structures on Iris dataset}. The permutation of CNOT gates leads to significant improvement in terms of validation loss. Moreover, extending the number of CNOT gates may reduce the performance of the embedding.}
    \label{fig:prelim-results}
\end{figure}
 \subsubsection{Encoding Scheme}
We proposed a representation for entangling layouts as directed multi-graphs, in which vertices represent qubits and edges formed by CNOT gates. The main objective of our proposed work is to find the optimal structure for entangling patterns on a given dataset. Hence, we fix the choice of rotation axis in the first and third parameterized unitary blocks at $\sigma_{1} = \sigma_{2}= Y$ \cite{lloyd2020quantum}. In other words, we only consider parameterized control rotation by Y-axis over all qubits. Moreover, the encoded graph representation for each candidate circuit layout in the search space is associated with an asymmetric adjacency matrix (Figure~\ref{fig:graph}), which allows better illustration in the complexity analysis.

\subsubsection{Complexity Analysis}
Given a set of $N$ qubits, the number of off-diagonal entries for the adjacent matrix is given as
\begin{equation}
    E = N^2-N = N(N-1)
\end{equation}
With regard the order permutation of CNOT gates in candidate circuit layouts, the total number of possible candidate in the search space of $N$ qubits embeddings is
\begin{equation}
    |\Omega_{\text{Full}}| = \sum_{k=0}^{E} k! \binom{E}{k} = \sum_{k=0}^{E}\frac{E!}{(E-k)!}.
\end{equation}
As a result, the search space of all possible circuit layouts is extensively large $(\approx1.3\times 10^9)$ even though we only consider a small system of $4$ qubits. Moreover, the complexity of the search space is exponentially expanded when increasing the number of input qubits, which is tremendously hard to find the optimal circuit layout for the entanglement block. Thus, we proposed additional parameters for the search space configuration, called the entanglement level, which is set to be the fixed number of CNOT gates within the entangling layer. Given a pre-defined entangling level $k$, the total number of possible circuit candidates in the reduced search space is:
\begin{equation}
    |\Omega_{\text{Reduced}}| = k! \binom{E}{k} = \frac{E!}{(E-k)!}.
\end{equation}
By implementing the proposed entangling level, the search space is reduced to a reasonable cardinality for finding optimal circuit architecture. Together with proposed encoding scheme (Figure~\ref{fig:encoding}), each circuit candidate in the search space of $N$ qubits constrained by entangling lever $k$ is represent by an encoded genotype vector $\bm{\alpha}$ of length $k$, whose each component is corresponding with each element of the ordered set of all possible connections $\mathcal{E} = \{e_i \}_{i \in \{ 0,...,E \}}$.

\subsection{Sequential Model-based Optimization}\label{smbo}
\begin{figure}
    \centering
    \includegraphics[width = 0.48\textwidth]{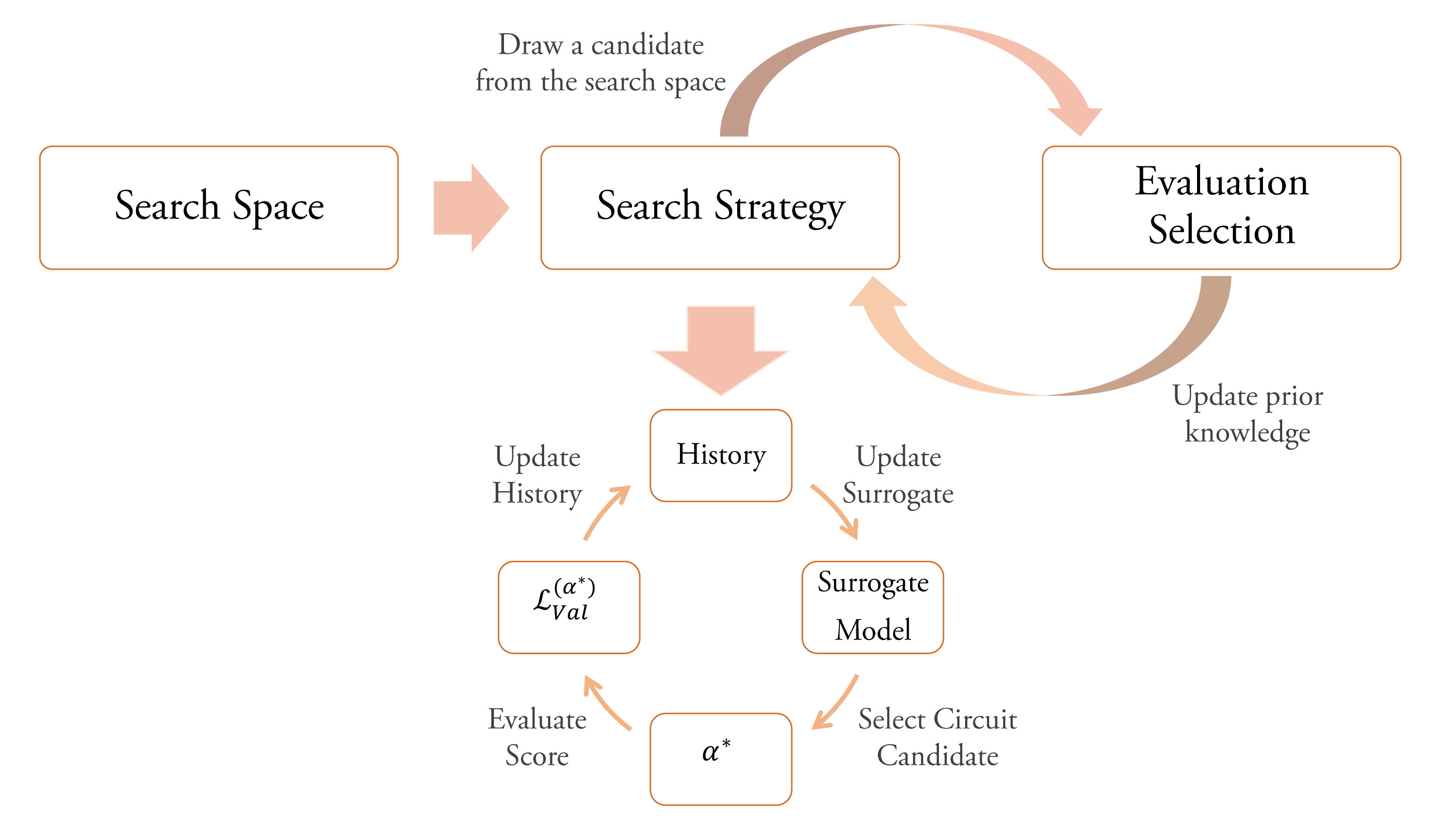}
    \caption{\textbf{General framework of automated model search}. After well-defining the search space, the automated intelligence draws an architecture candidate following search strategy. The drawn architecture will be evaluated using selection criteria, which create response for updating the search intelligence. In investigating search strategy SMBO-TPE, the prior knowledge of search intelligence will be updated corresponding to response score during the search phase.}
    \label{fig:smbo}
\end{figure}

\color{black}
Figure~\ref{fig:smbo} illustrates the framework of automated search for optimal configurations of the entanglement design. This procedure is motivated by hyper-parameters optimization, which has been well-studied in classical machine learning but not being utilized in quantum machine learning. For example, the optimization method is commonly used to find the optimal configurations of neural network (number of layers, initial channels or kernel size) or optimal training setting (learning rate or type of optimizer). It is worth mentioning that sequential model-based optimization (SMBO) enables efficient searching on highly complex search space (hundreds of dimensions\cite{bergstra2011algorithms}), which is a promising search strategy for the defined search space in Section~\ref{search-space}. Specifically, SMBO does not requires the computation of the true fitness function (true loss) to proposing the most potential candidates for the next trial. Instead, SMBO leverages surrogate models to approximate the true cost function and samples the most promising candidates based on selection criteria such as conditional entropy of minimizer, bandit-based criterion or expected improvement\cite{bergstra2011algorithms}. In this proposed research, we consider the expected improvement (EI) as the criterion for the sampler due to its versatility in different hyper-parameter settings\cite{bergstra2011algorithms}. This characteristic is not appeared in other search strategy such as reinforcement learning or gradient-based search since the methods demand the computation of the true cost function to update their samplers. 
\color{black}

We employ the sequential model-based optimization (SMBO) using tree Parzen estimator\cite{bergstra2011algorithms} as the search strategy for the optimal circuit candidate, which can be mathematically stated as
\begin{equation}
    \bm{\alpha^{*}} = \arg \min_{\bm{\alpha}\in \Omega} \mathcal{L}_{\text{val}}^{(\bm{\alpha})} (\bm{y},\hat{\bm{y}}),
\end{equation}
where $\hat{\bm{y}}$ is the prediction of the model. The cost function $\mathcal{L}_{\text{val}}^{(\bm{\alpha})}(\bm{y},\hat{\bm{y}})$ is the validation loss value corresponding to the circuit candidate $\bm{\alpha}$, which is modeled by less computational expensive surrogate model $S(\bm{\alpha})$ by SMBO.  \color{black}In other words, we approximate the true loss function by surrogates that are much simpler to be computed. Essentially, the usage of surrogates enables sampler to propose the most promising candidate for the next trial based on the prior knowledge. In the inner loop of Algorithm~\ref{algo:smbo}, we aim to optimize the Expected Improvement (EI) under surrogates $S(\bm{\alpha})$ using numerical optimization \color{black}. The Expected Improvement (EI) is the expected value from a given statistical model $f(x)$ in the model space $\mathcal{M}$ that the value $f(x)$ will be greater than a given threshold $\bar{t}$\color{black}. Mathematically, we have:
\begin{equation}\label{EI}
    \text{EI}_{\bar{t}} = \int_{-\infty}^{\infty} \max(\bar{t}-t)p_{M}(t|\bm{\alpha}) dt,
\end{equation}
where $M$ is an arbitrary model in the model space that $\mathcal{L}_{\text{val}}^{(\bm{\alpha})}(\bm{y},\hat{\bm{y}})$ will exceed $\bar{t}$. \color{black} Moreover, $p_M(t|\bm{\alpha})$s the conditional probability density of outputs corresponding to the candidate model parameterized by genotype vector $\bm{\alpha}$ \color{black}. The TPE estimator enables the decomposition of the conditional probability $p(t|\bm{\alpha})$ as two densities:
\begin{equation}
    p(\bm{\alpha}|t) = \begin{cases} l(\bm{\alpha}) \mbox{ if } t< \bar{t} \\
    g(\bm{\alpha}) \mbox{ if } t \geq \bar{t}
    \end{cases},
\end{equation}
reforming the EI in Equation~\ref{EI} into
\begin{equation}
    \begin{split}
    \text{EI}_{\bar{t}} &= \int_{-\infty}^{\bar{t}} (\bar{t}-t)\frac{p(\bm{\alpha}|t)p(t)}{p(\bm{\alpha})} dt \\
    &\propto \bigg[ p(t < \bar{t}) + \frac{g(\bm{\alpha})}{l(\bm{\alpha})}[1 - p(t < \bar{t})] \bigg]^{-1}
    \end{split}
\end{equation}
Due to the decomposition, the TPE estimator enables sampling multiple candidates based on the density $l(.)$, allowing more efficient estimation of the EI. \color{black}Furthermore, TPE estimators is different from Gaussian-based estimators (GP) in the choice of thresholding value for output $\bar{t}$. Specifically, GP favours values less than the best evaluation in the history while TPE favours $\bar{t}$ larger than such a best observation. Hence, TPE estimators can propose a threshold value correspond to some quantile of outputs, which enables more efficient sampling. Besides, SMBO-TPE is initialized with the prior distribution of discrete variables $p^{\text{prior}}_i$, which have the same length as the genotype vector. Thus, the posterior is proportional to $Lp^{\text{prior}}_i + C_{i}$, where $L$ is the length of genotype vector and $C_{i}$ is the number of choices for each element in the genotype. As a result, the search time of each trial using SMBO-TPE can be linearly scaled with the length of genotype vector with sorted observation in the history $\mathcal{H}$. This properties is very desirable since the search space of entangling layout is exponentially expanded with increasing number of input qubits or primitive gates. The final evaluation metric to compare the performance of derived architectures is the validation loss, which is computed by the validation dataset, illustrating the generalization of such embeddings. \color{black}

\color{black}

\begin{algorithm}[t]
\caption{Sequential Model-based Optimization via TPE estimator}
\label{algo:smbo}
\color{black}
Given search space $\Omega$ initialized by $k$, cost function $\mathcal{L}(.)$, initial model candidate $M_0$, $T$ number of iterations, surrogates $S(.)$ and history $\mathcal{H}$.\color{black}
\begin{enumerate}
    \item[] \textbf{initialize} $\mathcal{H} \leftarrow \emptyset$
    \item[] \textbf{for} $i=1$ to number of trials $T$:
    \begin{itemize}
        \item[] $\bm{\alpha}^{*} \leftarrow \arg \min_{\bm{\alpha}}S(\bm{\alpha},M_{i-1})$
        \item[] \textbf{compute}  $\mathcal{L}_{\text{val}}^{(\bm{\alpha^{*}})} (\bm{y},\hat{\bm{y}})$
        \item[] \textbf{update}  $\mathcal{H} \leftarrow \mathcal{H} \cup \{ \bm{\alpha^{*}} , \mathcal{L}_{\text{val}}^{(\bm{\alpha^{*}})} (\bm{y},\hat{\bm{y}})\} $
        \item[] \textbf{fit}  $M_i$ to $\mathcal{H}$
    \end{itemize}
    \item[] \textbf{return} $\mathcal{H}$
\end{enumerate}

\end{algorithm}

\color{black}
\subsection{Procedure}\label{procedure}
\begin{figure}[t]
    \centering
    \includegraphics[width = 0.48\textwidth]{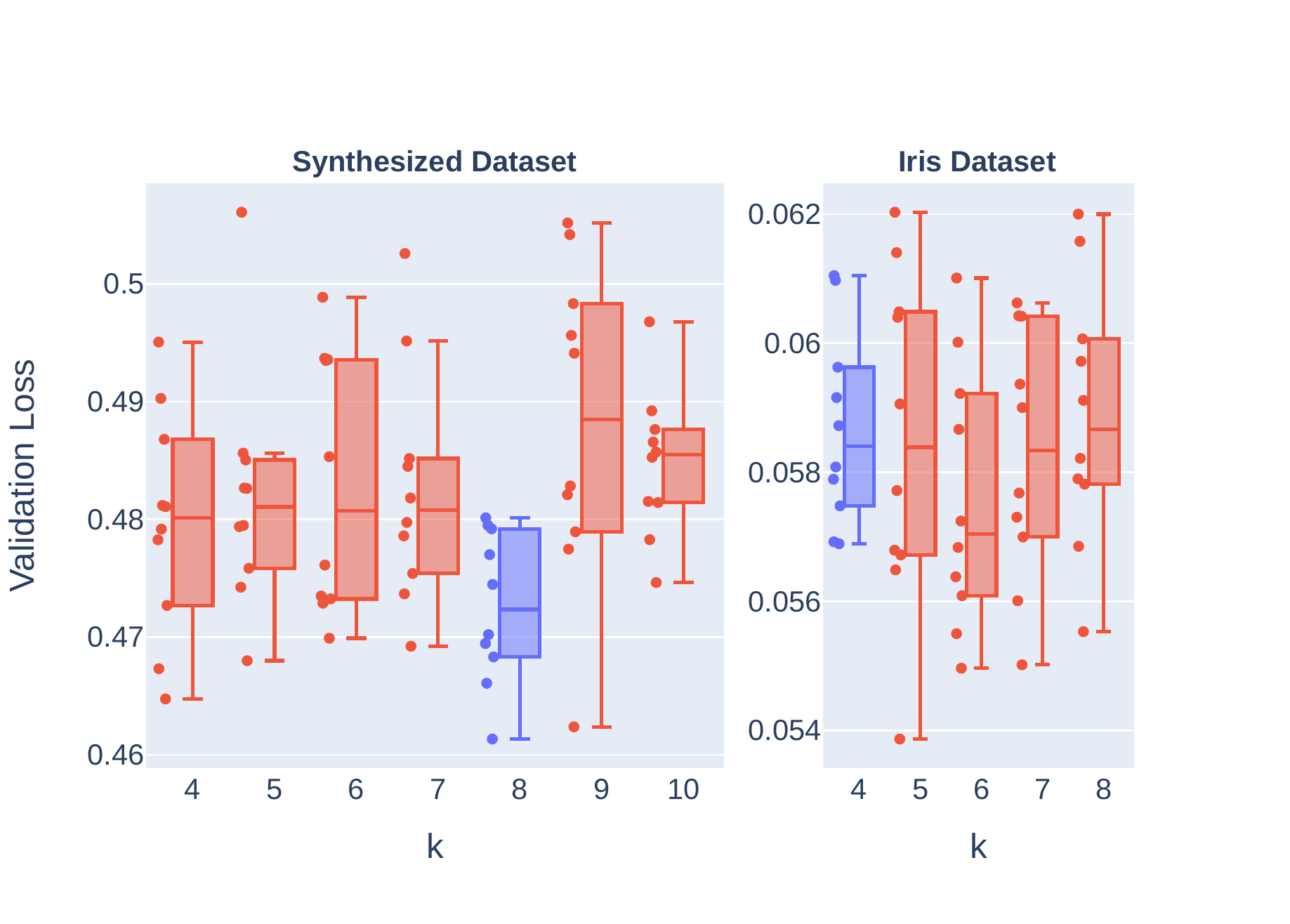}
    \caption{\textbf{Results from search phase with increasing number of entanglement levels $k$ on low dimensional datasets including synthesized and Iris datasets.} The best validation score on synthesized data is $0.4746\pm0.011$ using $k=8$. On the other hand, there is not enough statistical evidence to show that larger $k$ leads to better entanglement layout on Iris dataset.}
    \label{fig:k_values}
\end{figure}
We would like to summary the general procedure to implement our proposed framework in this section.\\
\textbf{Step 1 - Initialization:} The current implementation of quantum computers and quantum simulator only supports very limited number of noisy qubits, which restricts the capability of quantum embeddings within low-dimensional datasets. However, we can mitigate the curse of dimension for quantum embeddings by adopting hybrid classical-quantum neural architectures\cite{killoran2019continuous}. In these hybrid architectures, the classical component plays a role as an autoencoder (feature extractor), that transforms the input space $\mathbb{R}^{p} \rightarrow \mathbb{R}^{q}$, where $q < p$. Thus, if the given dataset is low-dimensional (number of features is less than number of available qubits), the original features will be directly used as input of quantum ansatz. Otherwise, classical autoencoder will be used to reduce the number of input features, then the resulting feature maps will be used as input of quantum embeddings. We will give more details about the hybrid classical-quantum architecture with use-cases in Section~\ref{experiments}.\\
\textbf{Step 2 - Search Phase:} Given $q$ number of input features, we create an ansatz (illustrated in Figure~\ref{fig:qe-qml}) with $q$ number of qubits for the quantum embeddings. Then, we initialize the first search space in Section~\ref{search-space} by selecting $k=q$ for the first iteration. This initial guess for $k$ is equivalent to baseline manual entanglement layouts such as fully entanglement (Figure~\ref{fig:graph}). We further perform searching by using increasing values for $k$ until the improvement gain appears diminishing. As a consequence, our proposed framework only guarantees to find the local optimal solution of the entanglement layout since we only investigate limited number of search spaces associated with the entanglement level. However, finding the global solutions remains an extremely difficult challenge for architecture optimization in machine learning. Therefore, the greedy heuristic for selecting $k$ used in our proposed work enables locally optimal solution while requiring reasonable computational resources such as hardware requirement or searching time.\\
\textbf{Step 3 - Evaluation Phase:} After discovering the entanglement structure, we compare the derived quantum ansatz with other classical counterparts using predictive performance as the evaluation metrics.

\section{Numerical Experiments}\label{experiments}

In this section, we report the experiment justifications for the effectiveness of our proposed QES by using different data scenarios. Incorporating with Section~\ref{procedure}, we first evaluate QES on low dimensional datasets using stand-alone quantum embeddings, including synthesized datasets and the IRIS dataset. The generated dataset includes $400$ observations in the feature space of dimension $4$, which involves the classification of three classes. We employ a small factor of hypercube size to obtain a synthesis dataset that is hard to be separated. Secondly, we further investigate the validity of QES on more challenging datasets, including Breast Cancer and Digits datasets. Although these datasets are not highly complex for classical machine learning, they can be considered high dimensional datasets for the current implementation of quantum computers and simulation. Hence, we leverage the hybrid classical-quantum neural architecture for these experiments. The detailed hyper-parameter setting for training the quantum embeddings is given in Appendix~\ref{apd:exp}. 
\color{black}

\subsection{Experimental Results of Stand-alone Quantum Embeddings on Low Dimensional Datasets}
\subsubsection{Synthesized Dataset}

In the search phase on synthesis dataset, we consider search spaces corresponding with increasing values of $k$. Moreover, the depth of each architecture candidate is set equal to two layers during both search phase and evaluation phase.

Figure~\ref{fig:found_circuit} illustrates the found architectures using SMBO-TPE and baseline random search. It is worth mentioning that the model complexity of the two setting is equivalent, that number of parameters in derived embedding architectures are equal. It is because the primitive CNOT gates does not contain any learnable weights. The score function for both of search procedure is based on the loss value on the (independent) validation set. We compare our found architectures with two common baseline entangling structures (strongly entangled layers) in Figure~\ref{fig:graph} and also with classical machine learning counterparts. In Table~\ref{tab:results}, our discovered quantum embedding circuit under gains a significant improvement in comparison to the baseline structures, while nearly achieve the performance of SVM and XGBoost with only minimal gap of $0.5\%$. Moreover, an obvious improvement of over $10\%$ gain in validation accuracy is witnessed in comparison with neural network with the same number of parameters. Besides, expanding the search space yields the locally optimal value $k=8$, which enables discovering circuit structures with higher predictive power, consistently for both search strategies. 

We further analyze the effectiveness of SMBO-TPE compared to baseline random search. Figure~\ref{fig:inter-results} shows the intermediate values of search strategy over trials of the two investigating optimization approaches. Overall, the validation loss converges after $50$ epochs for both settings of the search space. Moreover, optimal architectures are found in early trials using SMBO-TPE, while random search discovers such architectures in late trials of the search phase. Another advantage of SMBO-TPE over random search is depicted in the parallel coordinate plot, where we can see that the TPE sampler leverages the knowledge during the search phase to update its prior knowledge. Sampling results from TPE concentrate on edges that potentially form higher predictive performance, while the random search's sampled edges are widely spread throughout the configuration space. 

Finally, the search cost of quantum embeddings is significantly higher than searching for classical neural network architectures due to the computational limitation of near-term quantum simulators. For example, candidates in neural architecture search are convolution neural networks involving up to millions of parameters, which can be found in only $0.25-8$ GPU days by recent state-of-the-art NAS algorithms\cite{liu2018darts,nguyen2021contrastive}. On the other hand, training a minor quantum embedding of very few qubits consumes much larger computational expenses. Our experimental setting takes $2-4$ GPU days to search for an architecture of only $23$ parameters on the quantum simulators. Fortunately, the computational expenses are majorly accounted for training the quantum embeddings. In other words, the enhancement of quantum computing and quantum machine learning in the near-term devices that accelerates the trainability of quantum embeddings will be directly benefited by the proposed QES in term of search time.

\begin{figure}[t]
    \centering
    \includegraphics[width = 0.48\textwidth]{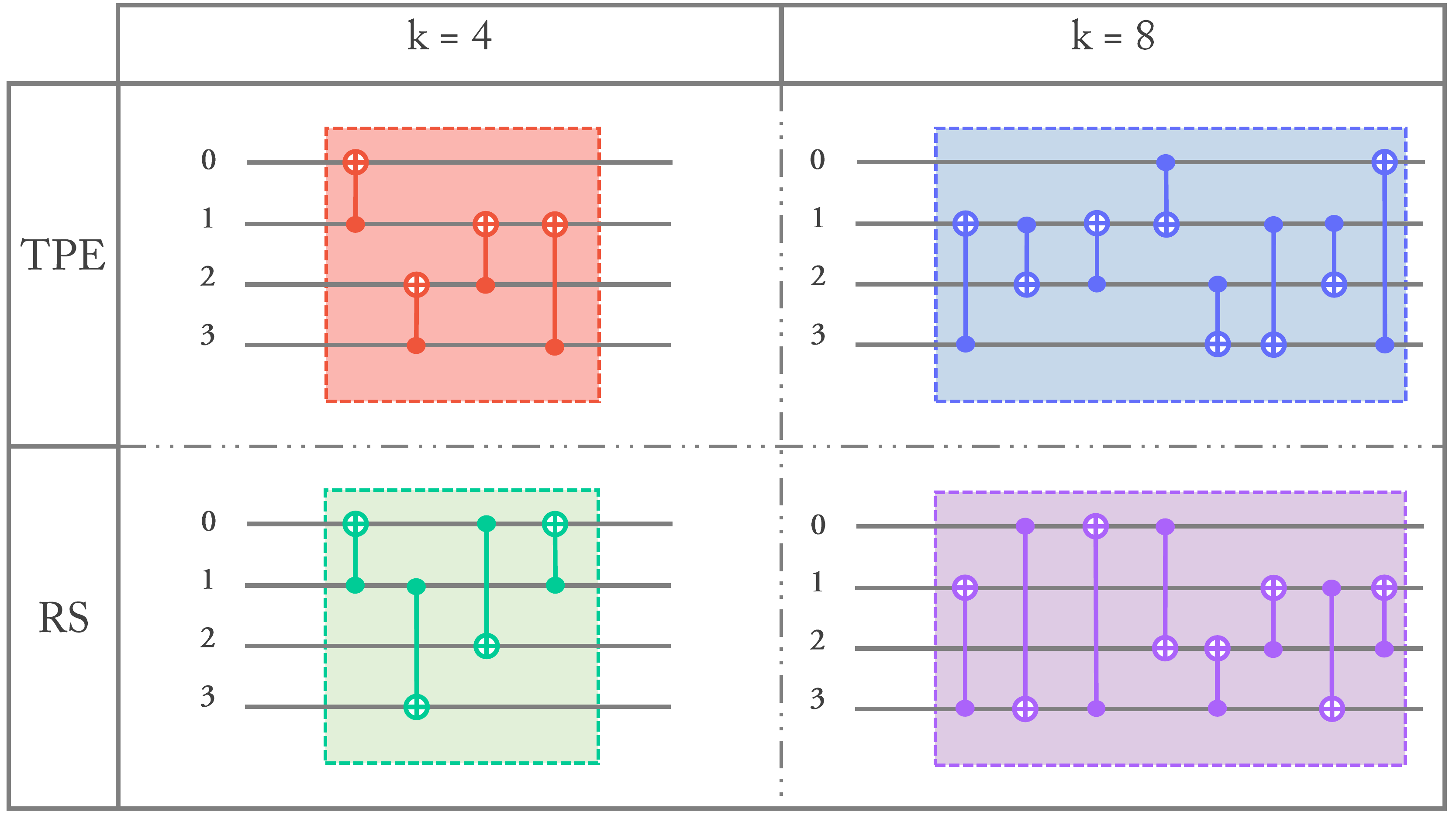}
    \caption{\textbf{Discovered quantum circuit architectures from different search space configurations and search strategies on synthesized dataset}. Found architecture contains a sequence of CNOT gates, which establishing entanglement over all qubits.}
    \label{fig:found_circuit}
\end{figure}
\begin{figure}[t]
    \centering
    \includegraphics[width = 0.48\textwidth]{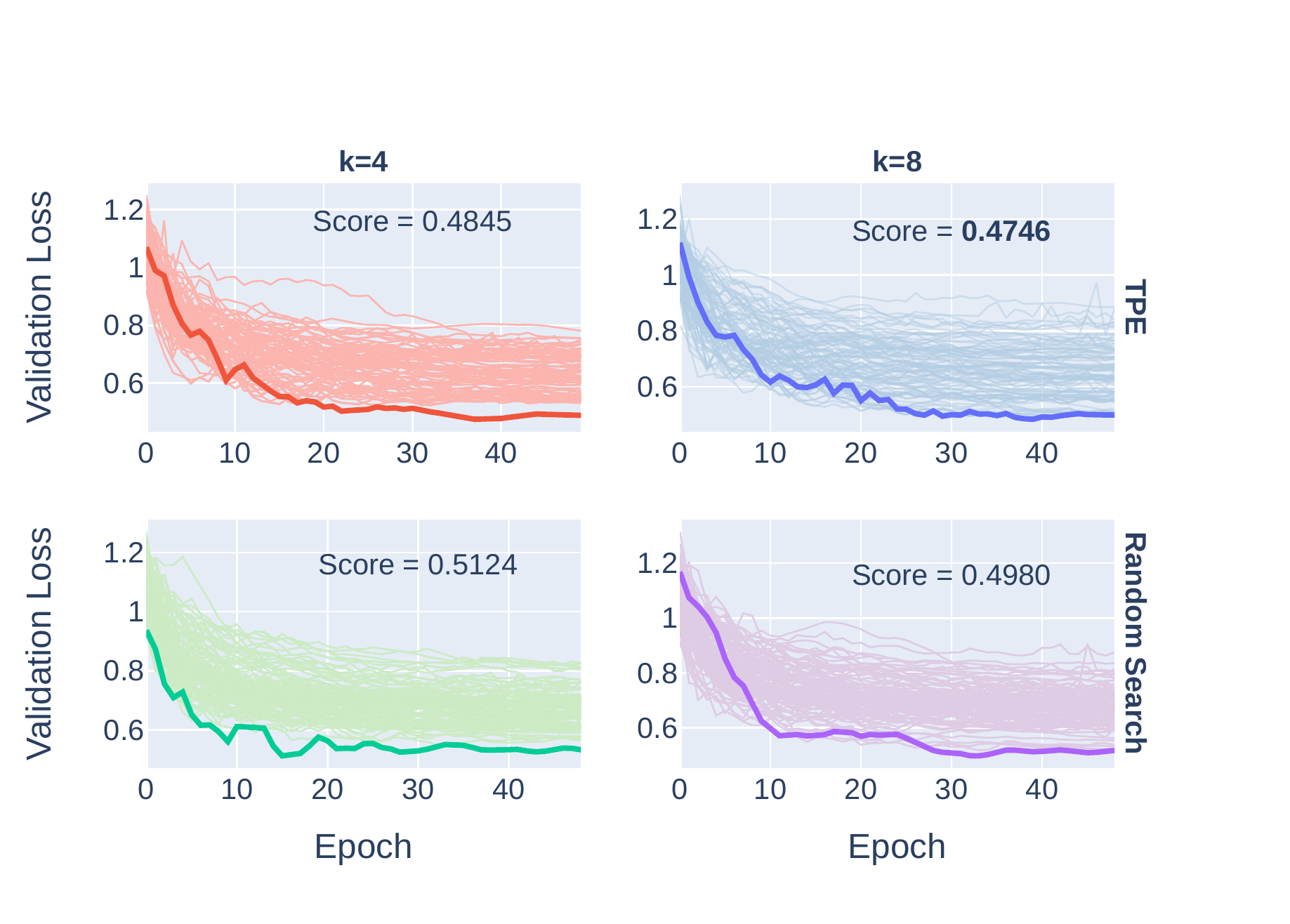}
    \caption{\textbf{Training response of all trials under each search space configurations and search strategies on synthesis dataset}. Each architecture candidate is trained for $100$ epochs at each trial. The validation loss converges within $50$ epochs (as shown). Best score is witnessed from the search space with $k=8$ under SBMO-TPE.}
    \label{fig:inter-results}
\end{figure}

\begin{table}[t]
    \centering
    \begin{tabular}{c|c|c}
        \toprule
         \textbf{Model} & \textbf{Accuracy} (\%) & \thead{\textbf{Search Cost} \\ \textbf{(GPU days)}}  \\
        \midrule
         Neural Network (fair) & $71.00$ & - \\
         SVM (ovo)\cite{bosertraining} & $84.00$ & -\\
         XGBoost\cite{chen2015xgboost} & $83.50$ & -\\
         \midrule
         Entangling Baseline 1 & $74.00$ & -\\
         Entangling Baseline 2 & $81.00$ & - \\
         \midrule
         QES-RS ($k=4$) & $79.00$ & $2$\\
         QES-RS ($k=8$) & $81.25$ & $4$ \\
         QES-TPE ($k=4$) & $82.00$ & $2$\\
         QES-TPE ($k=8$) & $83.50$ & $4$\\
         \bottomrule
    \end{tabular}
    \caption{\textbf{Comparison of found architectures with classical machine learning models and baseline hand-crafted entangling structure on synthesis dataset}. The evaluation is the validation accuracy based on 5 independent runs. Discovered quantum embedding outperforms baseline designs while achieves compatible performance as classical machine learning models.}
    \label{tab:results}
\end{table}
\begin{figure}[t]
    \centering
    \includegraphics[width = 0.24\textwidth]{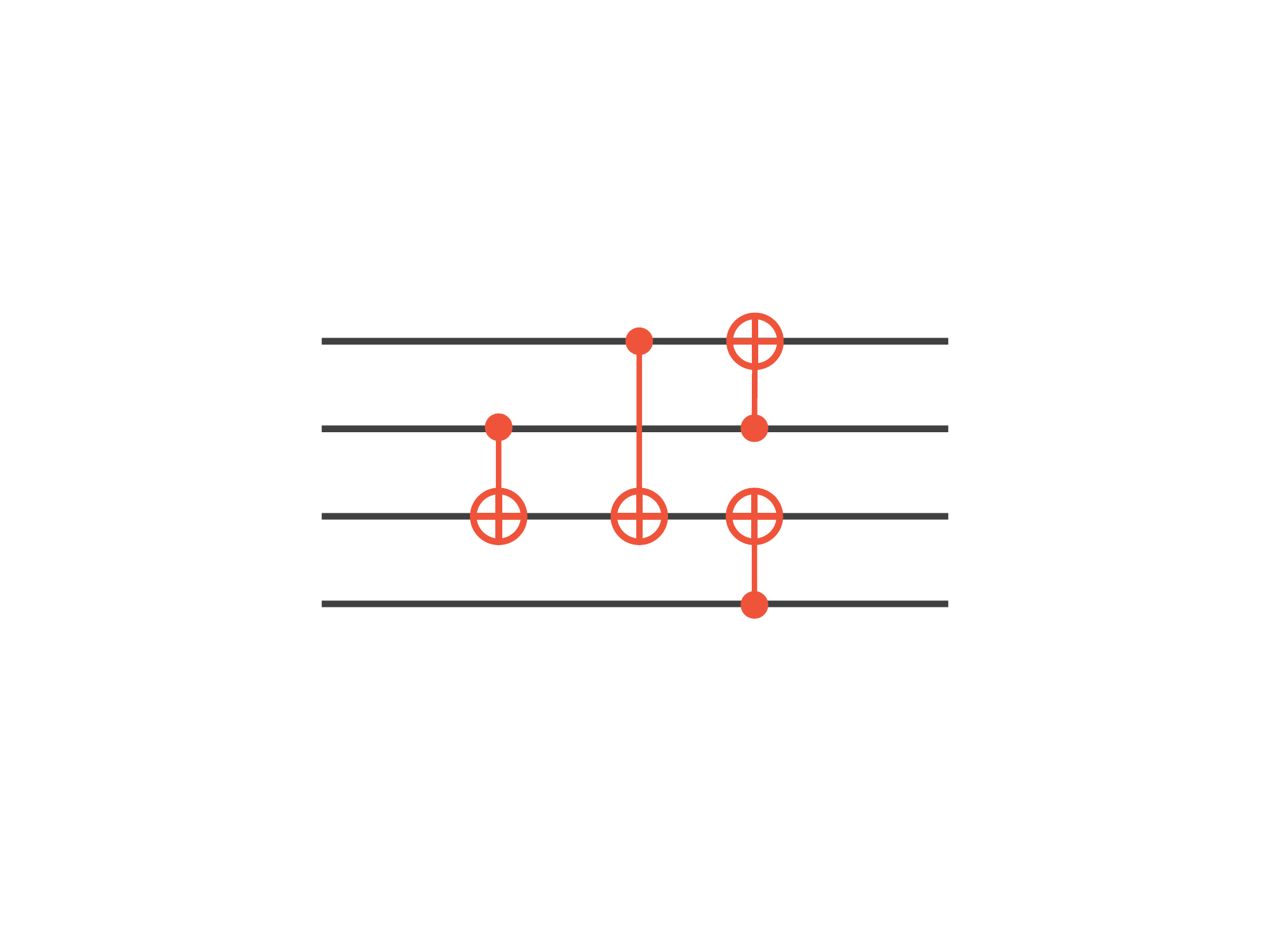}
    \includegraphics[width = 0.24\textwidth]{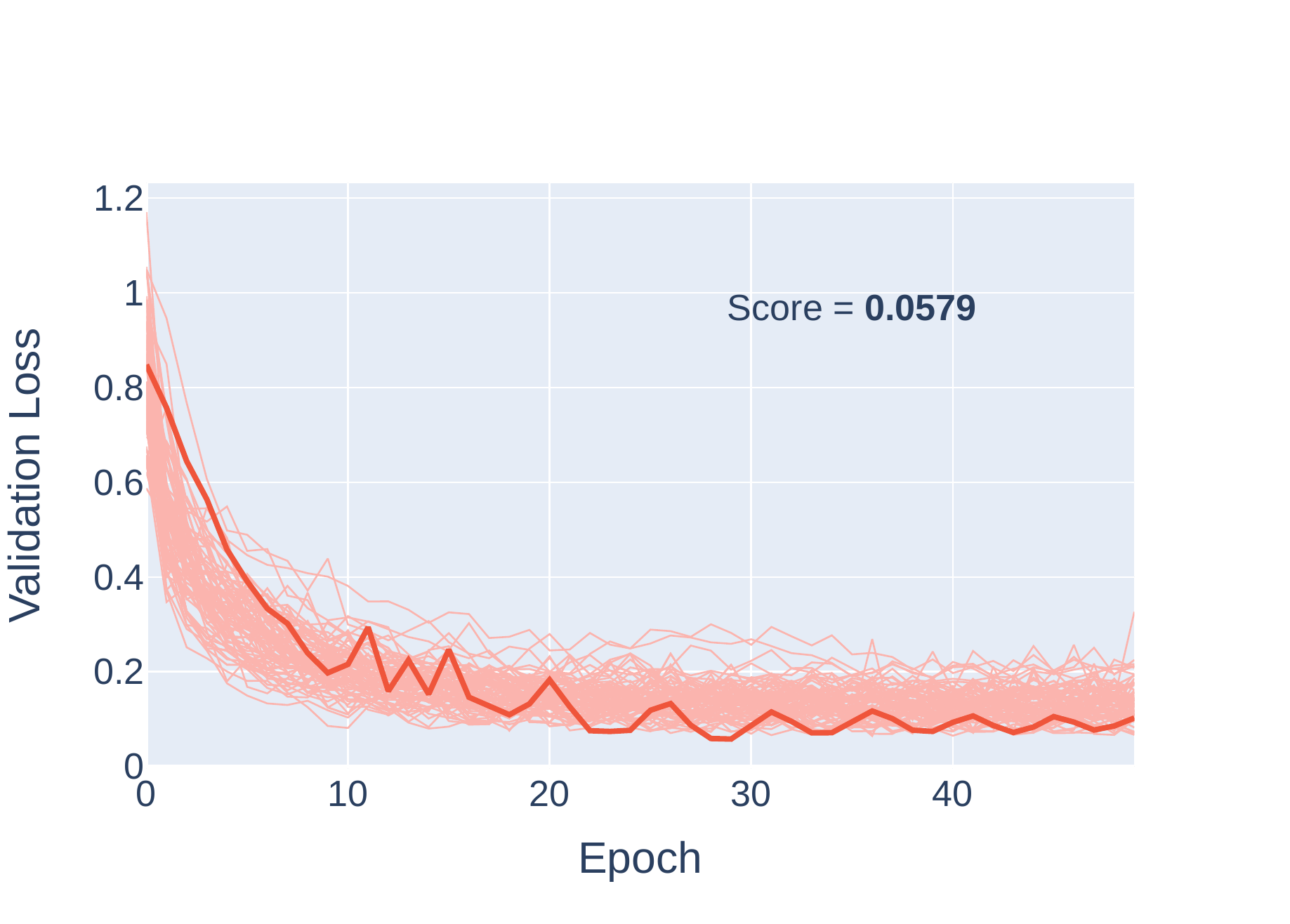}
    \caption{\textbf{Found architecture of quantum embedding and its optimization history}. The same pattern is as presented on simulated dataset. The full architecture of the quantum embedding includes stack of two identical layers with reported entanglement layout as in the left panel.}
    \label{fig:iris_results}
\end{figure}
\subsubsection{Results on Iris Dataset}
In this experiment, we use the original Iris datasets without any classical pre-processing, which includes 4 input features. We witness that the Iris dataset contains very well-representation observations, which is much easier to be separated in comparison to the synthesized dataset (Figure~\ref{fig:low-dim-data}). Hence, the magnitude of validation loss using different $k$ is hard to capture. From the left panel of Figure~\ref{fig:k_values}, there is not enough statistical evidence to show that increasing number of primitive results in better predictive performance. Thus, we only investigate the search space initialized by baseline $k=4$. 
\begin{table}[t]
    \centering
    \begin{tabular}{c|c|c}
    \toprule
       \textbf{\#Hidden Nodes} &  \textbf{\#Parameters} & \textbf{Test Accuracy}\\
        \midrule
        $6$ & $51$ & $93.00 \pm 0.0051$\\
        $7$ & $59$ &$95.00 \pm 0.0060$\\
        $8$ & $67$ & $96.67 \pm 0.0049$\\
        $9$ & $75$ & $96.67\pm 0.0049$\\
        $10$ & $83$ & $96.67\pm 0.0049$\\
        \midrule
        QES-RS & $23$ & $91.33\pm 0.0161$\\
         QES-TPE & $23$ & $95.33\pm 0.0125$\\
        \bottomrule
    \end{tabular}
    \caption{\textbf{Comparison between quantum and classical networks on Iris dataset.} The results is mean and standard deviation of test accuracy based on $100$ independent runs.}
    \label{tab:qe_vs_cnn}
\end{table}
We present the found structure for entanglement of quantum ansatz in Figure~\ref{fig:iris_results}, which involves a stack of two identical found architectures in the search phase. The same pattern in the simulated dataset, where the TPE sampler leverages the knowledge learned from response scores to update its prior distribution, enables better architectures. The discovered architecture achieves $95.33\pm0.0125$ in the validation accuracy (based on ten independent runs), outperforming two baseline designs close to $2\%$. Figure~\ref{fig:weights-iris} presents the convergence of model weights from founded quantum embedding, which indicates stable neural solution.

\color{black}
We further compare the proposed quantum embedding with classical embeddings using fair neural network. In the quantum classifier, there are $8$ learnable parameters for the variational rotation gates in the embeddings, which is followed by a classical post-processing using fully connected layer with $15$ parameters. Thus, the total number of parameter for the quantum classifier is $23$ parameters, including both quantum embedding and post-processing layer. We construct fair autoencoders which involves a single layer of hidden nodes (ranging from $6$ to $10$), followed by the same fully connected layer for the classifier. Since training small network may result in varying classification performances, we train each neural networks $100$ times and report their accuracy mean. As a result, the proposed quantum embedding outperforms a fairly classical neural network with $7$ hidden nodes by over $0.3\%$ ($95.33\%$ vs. $95\%$), while contains approximately $2.5\times$ less of parameters (Table~\ref{tab:qe_vs_cnn}).
\color{black}
\color{black}
\begin{figure}[t]
    \centering
    \includegraphics[width = 0.48\textwidth]{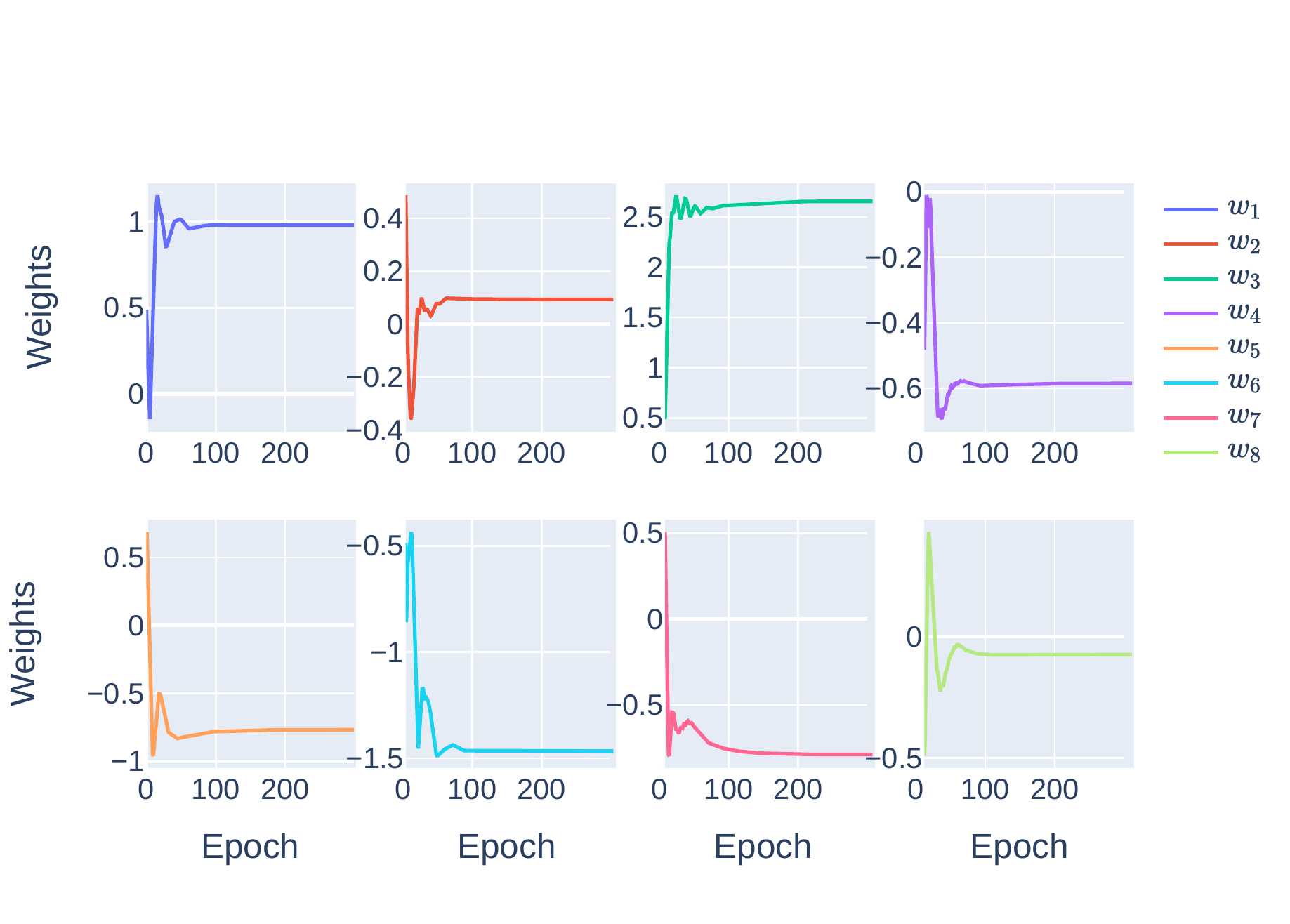}
    \caption{\textbf{Learning curves of quantum embedding's parameters on Iris dataset}. Model weights start converging after 100 epochs, achieve $95.33\pm0.0125$ in validation accuracy.}
    \label{fig:weights-iris}
\end{figure}
\subsection{Experimental Results of Hybrid Classical-Quantum on Higher Dimensional Datasets}
\begin{figure}[t]
    \centering
    \includegraphics[width = 0.48\textwidth]{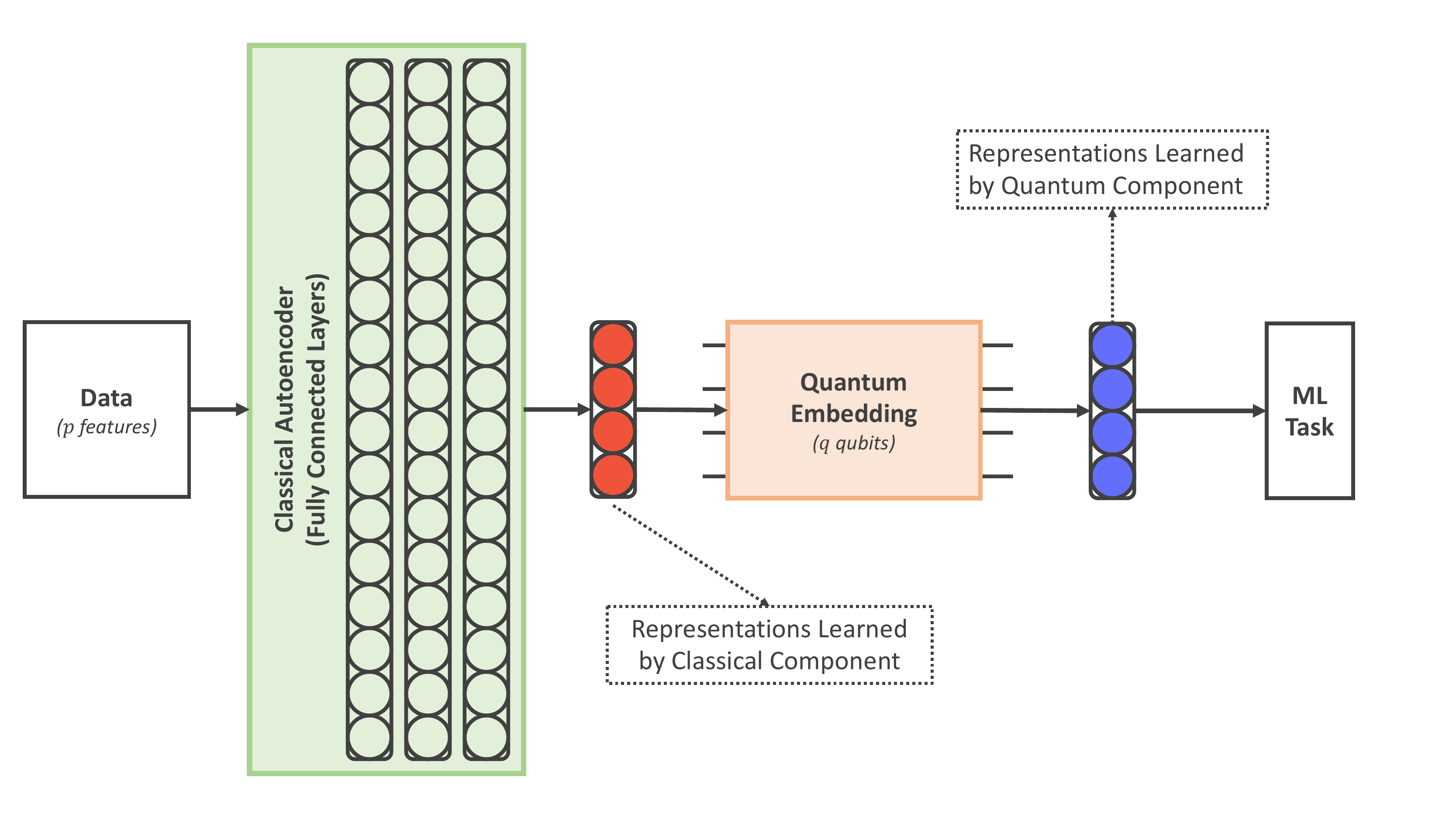}
    \caption{\textbf{Structure of Hybrid Classical-Quantum Neural Architecture}. The data scenarios is when the number of input features ($p$) is larger than the number of available qubits ($q$). We decompose the architecture to investigate the representation learning ability of each component.}
    \label{fig:hybrid-nn}
\end{figure}
\begin{figure}
    \centering
    \includegraphics[width = 0.48\textwidth]{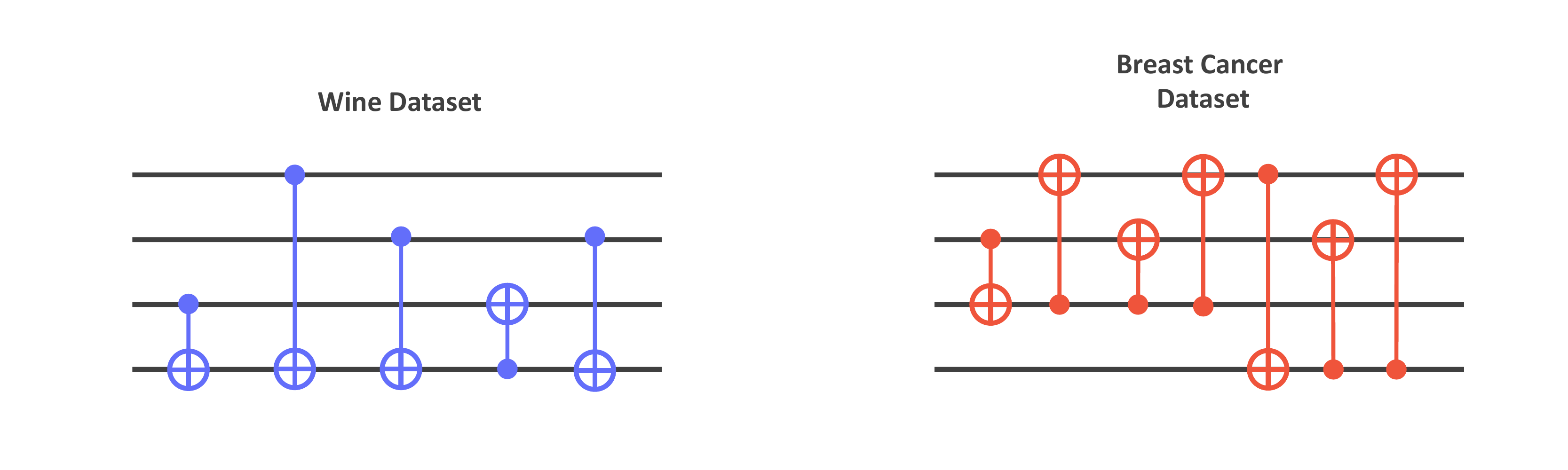}
    \caption{\textbf{Discovered entanglement layouts using hybrid classical-quantum neural architectures.} The final ansatz includes a stack of two layers.}
    \label{fig:found-circuit-high}
\end{figure}

\begin{figure*}
    \centering
    \includegraphics[width = 0.48\textwidth]{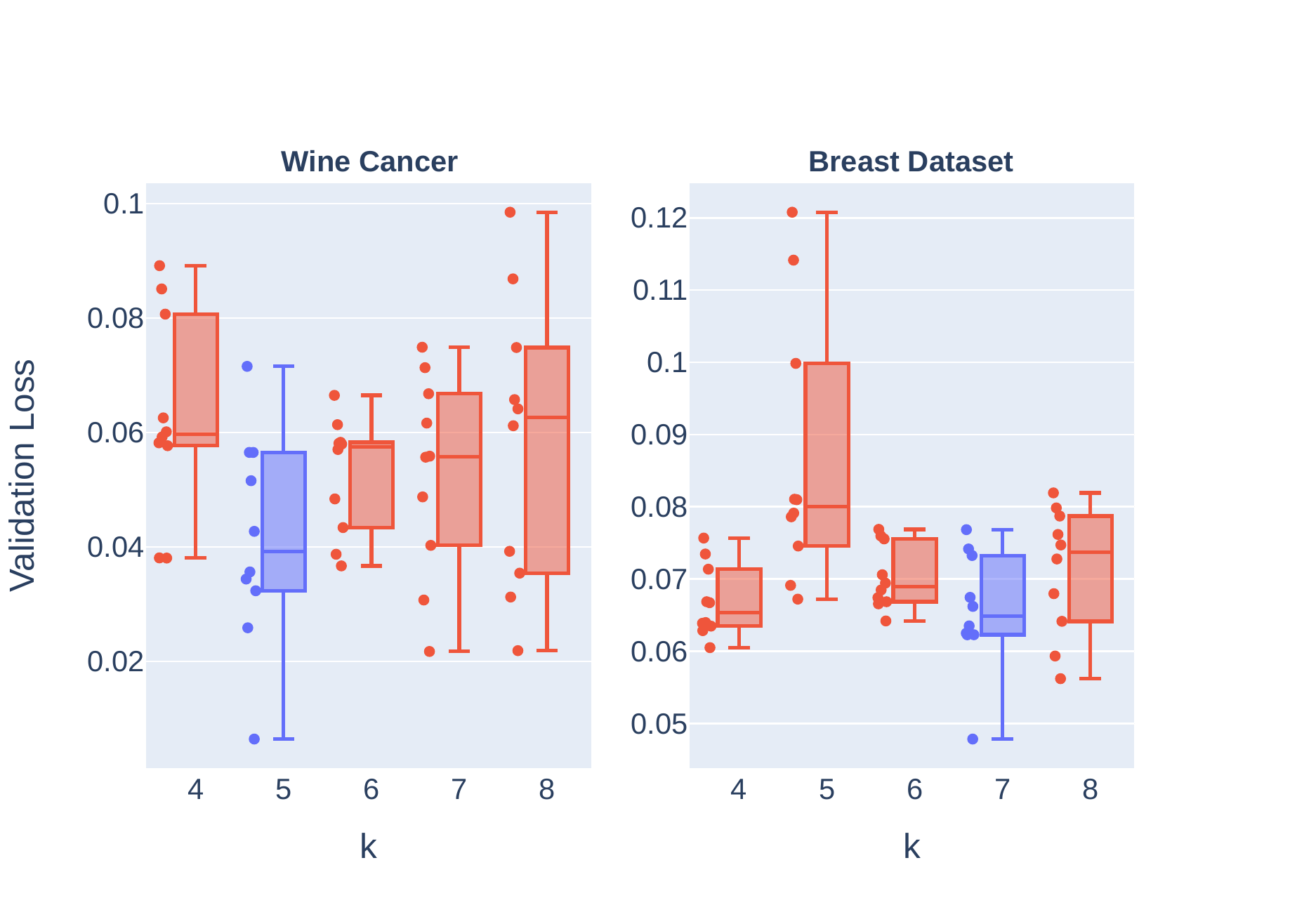}
    \includegraphics[width = 0.48\textwidth]{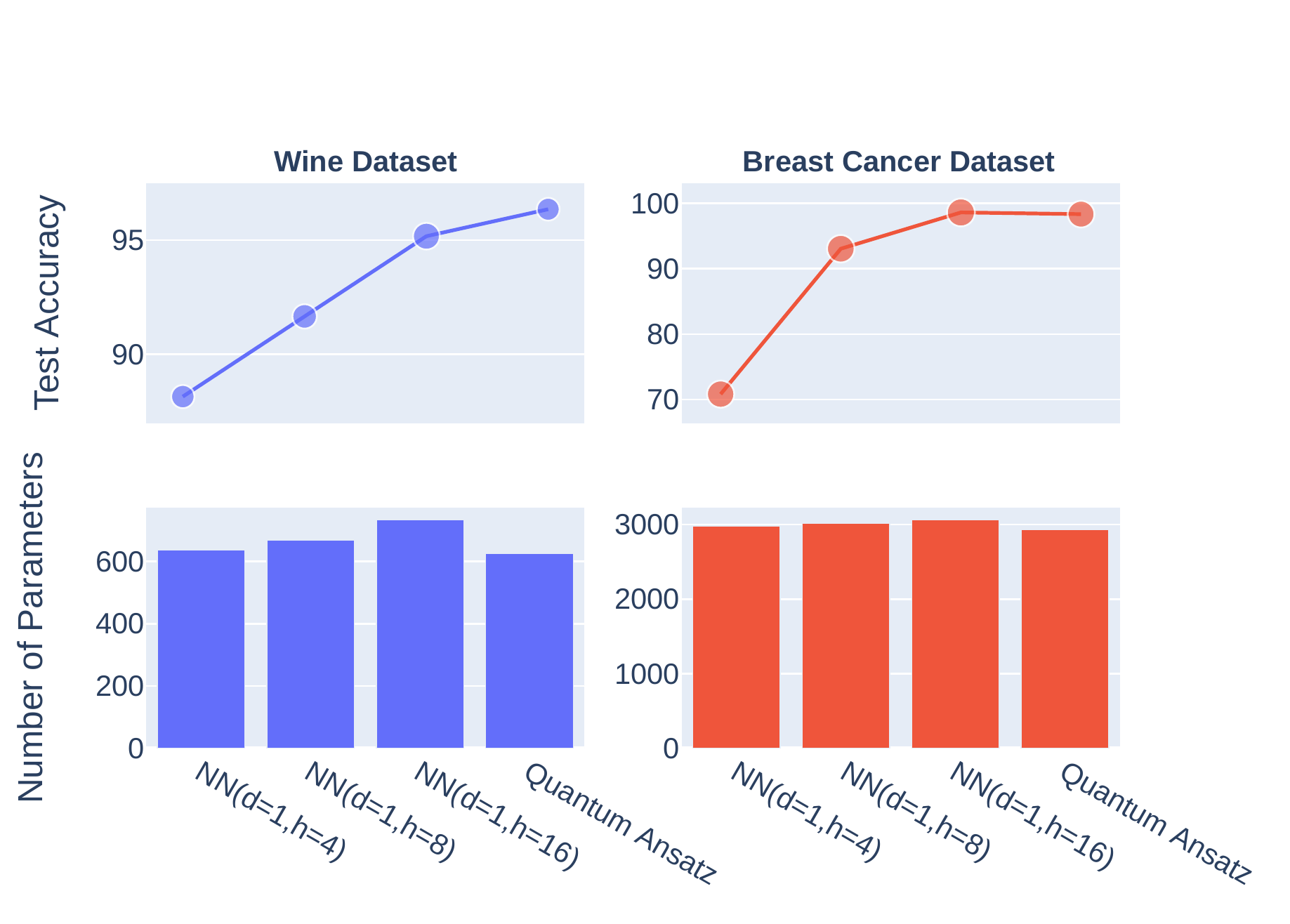}
    
    \caption{\textbf{Results from the search phase using hybrid classical-quantum neural architecture on breast cancer and wine dataset.} (Left)The locally optimal value for entangling level from the breast cancer data is $k=7$, while that from wine data is $k=5$. (Right) Evaluation results from found ansatz, in comparison to fair classical neural networks.}
    \label{fig:high-result}
\end{figure*}

\begin{figure}
    \centering
    \includegraphics[width = 0.48\textwidth]{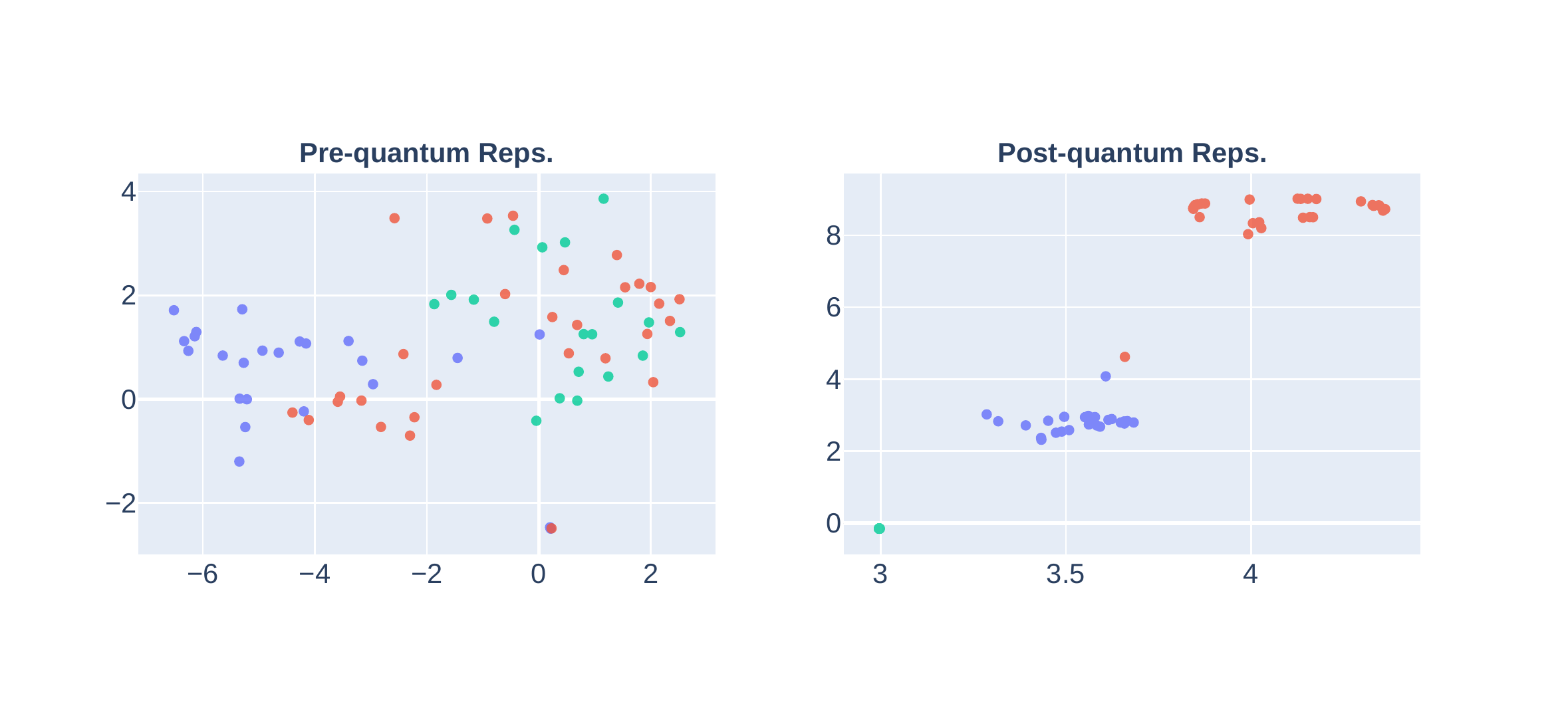}
    \includegraphics[width = 0.48\textwidth]{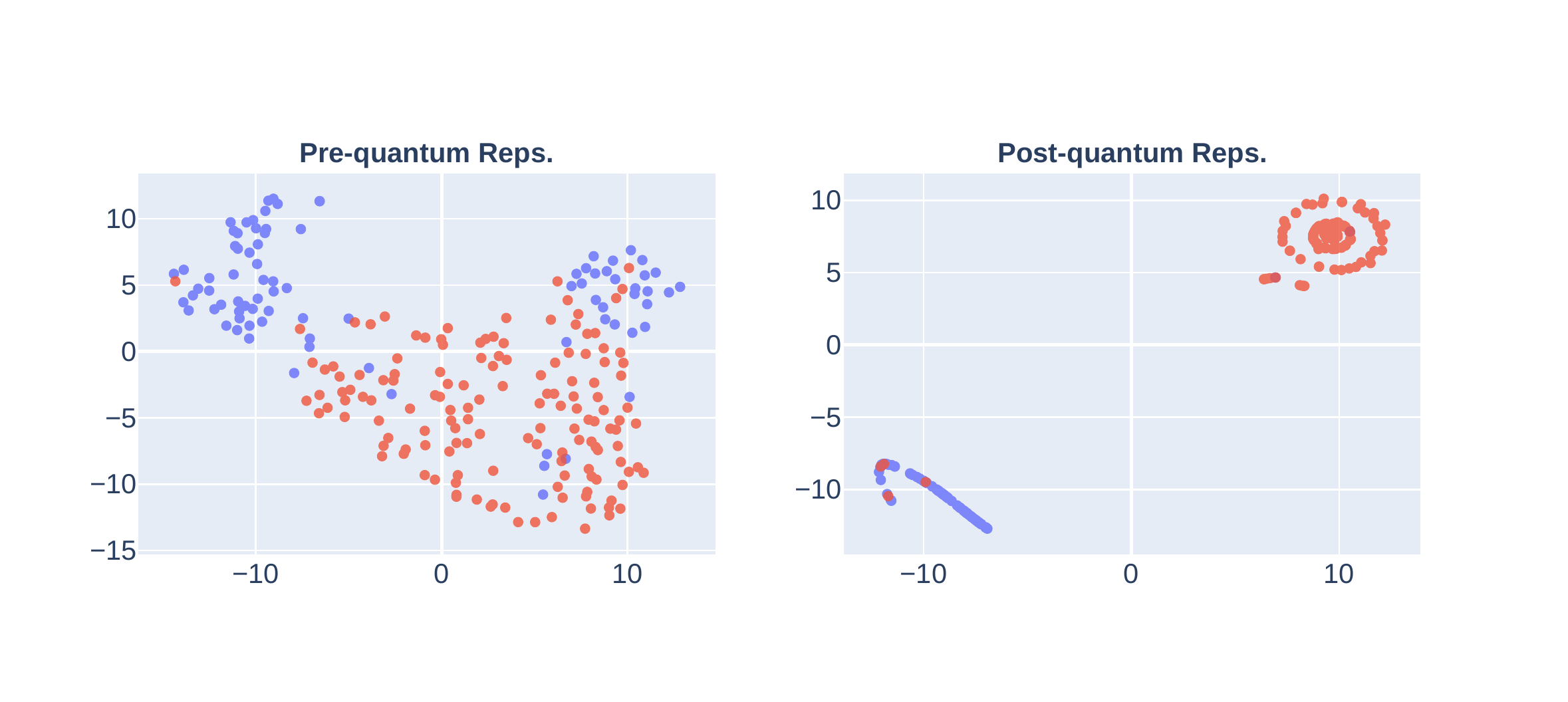}
    \caption{\textbf{Analysis on representation learning pre- and post- quantum embeddings using T-SNE: (Top) Wine dataset, (Bottom) Breast Cancer dataset.} The T-SNE features from pre-ansatz representations is hard to be separated using linear classifier, while those features from post-representations can be well-separated using the same classifiers.}
    \label{fig:reps}
\end{figure}
In this section, we evaluate the proposed QES on more challenging data scenarios. These datasets have higher number of original features than the number of available qubits on quantum ansatz, which is summarized in Table~\ref{tab:high-data}. We leverage a classical autoencoder, which includes a stack of $3$ fully connected layers, to reduce the dimensionality of input dataset for quantum ansatz (Figure~\ref{fig:hybrid-nn}). The usage of hybrid neural architectures may raise the concern whether the effectiveness of learning comes from classical or quantum components. Therefore, we also analyze the representation learning ability from each component by compare pre- and post-quantum feature maps. The detailed comparison will be given within each case study hereafter. 

\color{black}
The left panel of Figure~\ref{fig:high-result} shows the optimal values for entanglement level $k$ following procedure in Section~\ref{procedure}. On the wine dataset, the optimal quantum circuit is identified in the search space generated by $k=5$, which is visualized in the left panel of Figure~\ref{fig:found-circuit-high}. On the other hand, $k=7$ is found in breast cancer dataset, resulting in the structure of CNOT depicted in the right panel of Figure~\ref{fig:found-circuit-high}\color{black}. We analyze the performance of found ansatz and compare to fair classical neural networks. It is important to emphasize that we only replace the quantum asatz by a fully connected layer. In other words, the classical autoencoder (Figure~\ref{fig:hybrid-nn}) is used in all experiments and contributes a static number of parameters to the model complexity. We report the configuration of fair classical neural networks (NN) in the right panel of Figure~\ref{fig:high-result}, in which the NN has single depth $d=1$ with $h$ hidden nodes. In addition, the total number of weights of the whole hybrid architecture is reported for comparisons. As a result, the proposed quantum ansatz outperforms classical NN in the wine dataset (based on $100$ independent runs) while maintains less a number of parameters. Regarding Breast Cancer dataset, the performance of discovered ansatz is higher than NNs with $4$ and $8$ hidden nodes, but slightly lower than NN with $h = 16$ despite possesses the least number of parameters.

We aware that the usage of hybrid architectures may raise a critical concern about the effectiveness of found ansatz. Particularly, it is not hard to wonder that the learning has already done by the classical autoencoders, since they have a relatively large model complexity for such less complex problems like Wine and Breast Cancer datasets. Thus, we decompose the neural architectures and analyse the representations pre- and post- quantum ansatz (red and blue vectors in Figure~\ref{fig:hybrid-nn}). The evaluation of representation learning can be delivered using T-SNE\cite{van2008visualizing} visualization of feature vectors, in which we can observes the distance between clusters of classes, depicted in Figure~\ref{fig:reps}. In the T-SNE visualization from the wine dataset, the decision boundary between clusters are unclear when we visualize the pre-ansatz features. In contrast, T-SNE of post-ansatz features shows clearer boundary between classes, resulting in three separated cluster. This is consistent to the predictive performance of the hybrid architecture with ansatz, which possesses the test accuracy of $98.34 \pm 0.0021$. The same observations can be seen from Breast Cancer dataset, where the T-SNE of pre-quantum representations is difficult to be separated by linear classifier, while that from post-ansatz features are well-separated. The experiments enable insights into the effectiveness of found ansatz, which leads to more efficient representation learning, in term of predictive performance.

\color{black}
\section{Discussion and Conclusion}\label{discusion}
\subsection{Implication}
Beyond the numerical experiments, we would like to address several general principles from our QES. Our proposed approach provides an automated search intelligence that can find an optimal architecture of a quantum embedding circuit for a given dataset. It is reasonable to believe that there is no universal design of the embedding structure for every dataset, but instead, we can derive optimal architectures that well-performs on the dataset.

\subsection{Threads to Validity}
Threats to the internal validity of QES consider the reproducible ability of the algorithm, which is the most challenging factor in automated machine learning \cite{elsken2019neural,ren2020comprehensive}. As we are in the noisy intermediate-scale quantum era, such issues have been amplified compared to classical computational hardware. Moreover, our proposed QES relies on the assumption of the entangling level that tremendously reduces the cardinal of the search space. Hence, we lack empirical evidence of the effectiveness of QES on expanded search space, especially when the number of qubits is scaled. We want to defer the investigation of such a problem for further study. Nevertheless, QES well-performs in small qubits can achieve similar results with classical machine learning counterparts and outperform basic entangling structures.

Threats to external validity include the generalization of QES on different data scenarios, the number of qubits in the system, and noises inherited from actual quantum computing hardware. Although the experimental results from simulated quantum computing hardware are robust and stable, the story may change when we implemented QES on near-term noisy quantum computers. Another thread is the computational limitation of near-term quantum computers and quantum simulators. The cost for training single quantum embedding is remarkably higher than training classical encoders in the quantum simulations, which leads to a very high computational expense for the search phase. These threats indicate future research opportunities in quantum embeddings, including implementation of different search strategies.

\subsection{Conclusion}
This paper proposes an automated procedure for finding optimal quantum embeddings architecture that leads to high representation learning ability on the quantum Hilbert space. The algorithm is accessible and promising compared to the classical machine learning model, which can be implemented on near-term quantum computers. Although our QES cannot guarantee to find the global optimal design of the quantum embedding architecture in any full search space, it can certainly discover high-performed architecture solutions under the constrain of the entanglement level.

\bibliography{output}

\begin{thebibliography}{10}
\providecommand{\url}[1]{#1}
\csname url@samestyle\endcsname
\providecommand{\newblock}{\relax}
\providecommand{\bibinfo}[2]{#2}
\providecommand{\BIBentrySTDinterwordspacing}{\spaceskip=0pt\relax}
\providecommand{\BIBentryALTinterwordstretchfactor}{4}
\providecommand{\BIBentryALTinterwordspacing}{\spaceskip=\fontdimen2\font plus
\BIBentryALTinterwordstretchfactor\fontdimen3\font minus
  \fontdimen4\font\relax}
\providecommand{\BIBforeignlanguage}[2]{{%
\expandafter\ifx\csname l@#1\endcsname\relax
\typeout{** WARNING: IEEEtran.bst: No hyphenation pattern has been}%
\typeout{** loaded for the language `#1'. Using the pattern for}%
\typeout{** the default language instead.}%
\else
\language=\csname l@#1\endcsname
\fi
#2}}
\providecommand{\BIBdecl}{\relax}
\BIBdecl

\bibitem{huang2021power}
H.-Y. Huang, M.~Broughton, M.~Mohseni, R.~Babbush, S.~Boixo, H.~Neven, and
  J.~R. McClean, ``Power of data in quantum machine learning,'' \emph{Nature
  communications}, vol.~12, no.~1, pp. 1--9, 2021.

\bibitem{schuld2021supervised}
M.~Schuld, ``Supervised quantum machine learning models are kernel methods,''
  \emph{arXiv preprint arXiv:2101.11020}, 2021.

\bibitem{havlivcek2019supervised}
V.~Havl{\'\i}{\v{c}}ek, A.~D. C{\'o}rcoles, K.~Temme, A.~W. Harrow, A.~Kandala,
  J.~M. Chow, and J.~M. Gambetta, ``Supervised learning with quantum-enhanced
  feature spaces,'' \emph{Nature}, vol. 567, no. 7747, pp. 209--212, 2019.

\bibitem{schuld2019quantum}
M.~Schuld and N.~Killoran, ``Quantum machine learning in feature hilbert
  spaces,'' \emph{Physical review letters}, vol. 122, no.~4, p. 040504, 2019.

\bibitem{lloyd2020quantum}
S.~Lloyd, M.~Schuld, A.~Ijaz, J.~Izaac, and N.~Killoran, ``Quantum embeddings
  for machine learning,'' \emph{arXiv preprint arXiv:2001.03622}, 2020.

\bibitem{benedetti2019parameterized}
M.~Benedetti, E.~Lloyd, S.~Sack, and M.~Fiorentini, ``Parameterized quantum
  circuits as machine learning models,'' \emph{Quantum Science and Technology},
  vol.~4, no.~4, p. 043001, 2019.

\bibitem{farhi2018classification}
E.~Farhi and H.~Neven, ``Classification with quantum neural networks on near
  term processors,'' \emph{arXiv preprint arXiv:1802.06002}, 2018.

\bibitem{mcclean2018barren}
J.~R. McClean, S.~Boixo, V.~N. Smelyanskiy, R.~Babbush, and H.~Neven, ``Barren
  plateaus in quantum neural network training landscapes,'' \emph{Nature
  communications}, vol.~9, no.~1, pp. 1--6, 2018.

\bibitem{romero2021variational}
J.~Romero and A.~Aspuru-Guzik, ``Variational quantum generators: Generative
  adversarial quantum machine learning for continuous distributions,''
  \emph{Advanced Quantum Technologies}, vol.~4, no.~1, p. 2000003, 2021.

\bibitem{mcclean2016theory}
J.~R. McClean, J.~Romero, R.~Babbush, and A.~Aspuru-Guzik, ``The theory of
  variational hybrid quantum-classical algorithms,'' \emph{New Journal of
  Physics}, vol.~18, no.~2, p. 023023, 2016.

\bibitem{bergstra2011algorithms}
J.~Bergstra, R.~Bardenet, Y.~Bengio, and B.~K{\'e}gl, ``Algorithms for
  hyper-parameter optimization,'' vol.~24, 2011.

\bibitem{fisher1936use}
R.~A. Fisher, ``The use of multiple measurements in taxonomic problems,''
  \emph{Annals of eugenics}, vol.~7, no.~2, pp. 179--188, 1936.

\bibitem{zoph2016neural}
B.~Zoph and Q.~V. Le, ``Neural architecture search with reinforcement
  learning,'' \emph{arXiv preprint arXiv:1611.01578}, 2016.

\bibitem{zoph2018learning}
B.~Zoph, V.~Vasudevan, J.~Shlens, and Q.~V. Le, ``Learning transferable
  architectures for scalable image recognition,'' in \emph{Proceedings of the
  IEEE conference on computer vision and pattern recognition}, 2018, pp.
  8697--8710.

\bibitem{bergstra2012random}
J.~Bergstra and Y.~Bengio, ``Random search for hyper-parameter optimization.''
  \emph{Journal of machine learning research}, vol.~13, no.~2, 2012.

\bibitem{snoek2012practical}
J.~Snoek, H.~Larochelle, and R.~P. Adams, ``Practical bayesian optimization of
  machine learning algorithms,'' \emph{arXiv preprint arXiv:1206.2944}, 2012.

\bibitem{real2019regularized}
E.~Real, A.~Aggarwal, Y.~Huang, and Q.~V. Le, ``Regularized evolution for image
  classifier architecture search,'' in \emph{Proceedings of the aaai conference
  on artificial intelligence}, vol.~33, no.~01, 2019, pp. 4780--4789.

\bibitem{liu2018darts}
H.~Liu, K.~Simonyan, and Y.~Yang, ``Darts: Differentiable architecture
  search,'' \emph{arXiv preprint arXiv:1806.09055}, 2018.

\bibitem{liu2018progressive}
C.~Liu, B.~Zoph, M.~Neumann, J.~Shlens, W.~Hua, L.-J. Li, L.~Fei-Fei,
  A.~Yuille, J.~Huang, and K.~Murphy, ``Progressive neural architecture
  search,'' in \emph{Proceedings of the European conference on computer vision
  (ECCV)}, 2018, pp. 19--34.

\bibitem{nguyen2021contrastive}
N.~Nguyen and J.~M. Chang, ``Contrastive self-supervised neural architecture
  search,'' \emph{arXiv preprint arXiv:2102.10557}, 2021.

\bibitem{fosel2021quantum}
T.~F{\"o}sel, M.~Y. Niu, F.~Marquardt, and L.~Li, ``Quantum circuit
  optimization with deep reinforcement learning,'' \emph{arXiv preprint
  arXiv:2103.07585}, 2021.

\bibitem{liu2018quantum}
N.~Liu and P.~Rebentrost, ``Quantum machine learning for quantum anomaly
  detection,'' \emph{Physical Review A}, vol.~97, no.~4, p. 042315, 2018.

\bibitem{gao2018quantum}
X.~Gao, Z.-Y. Zhang, and L.-M. Duan, ``A quantum machine learning algorithm
  based on generative models,'' \emph{Science advances}, vol.~4, no.~12, p.
  eaat9004, 2018.

\bibitem{dunjko2016quantum}
V.~Dunjko, J.~M. Taylor, and H.~J. Briegel, ``Quantum-enhanced machine
  learning,'' \emph{Physical review letters}, vol. 117, no.~13, p. 130501,
  2016.

\bibitem{von2018quantum}
O.~A. Von~Lilienfeld, ``Quantum machine learning in chemical compound space,''
  \emph{Angewandte Chemie International Edition}, vol.~57, no.~16, pp.
  4164--4169, 2018.

\bibitem{lloyd2016quantum}
S.~Lloyd, S.~Garnerone, and P.~Zanardi, ``Quantum algorithms for topological
  and geometric analysis of data,'' \emph{Nature communications}, vol.~7,
  no.~1, pp. 1--7, 2016.

\bibitem{lloyd2014quantum}
S.~Lloyd, M.~Mohseni, and P.~Rebentrost, ``Quantum principal component
  analysis,'' \emph{Nature Physics}, vol.~10, no.~9, pp. 631--633, 2014.

\bibitem{farhi2014quantum}
E.~Farhi, J.~Goldstone, and S.~Gutmann, ``A quantum approximate optimization
  algorithm,'' \emph{arXiv preprint arXiv:1411.4028}, 2014.

\bibitem{mari2020transfer}
A.~Mari, T.~R. Bromley, J.~Izaac, M.~Schuld, and N.~Killoran, ``Transfer
  learning in hybrid classical-quantum neural networks,'' \emph{Quantum},
  vol.~4, p. 340, 2020.

\bibitem{cappelletti2020polyadic}
W.~Cappelletti, R.~Erbanni, and J.~Keller, ``Polyadic quantum classifier,'' in
  \emph{2020 IEEE International Conference on Quantum Computing and Engineering
  (QCE)}.\hskip 1em plus 0.5em minus 0.4em\relax IEEE, 2020, pp. 22--29.

\bibitem{schuld2020circuit}
M.~Schuld, A.~Bocharov, K.~M. Svore, and N.~Wiebe, ``Circuit-centric quantum
  classifiers,'' \emph{Physical Review A}, vol. 101, no.~3, p. 032308, 2020.

\bibitem{killoran2019continuous}
N.~Killoran, T.~R. Bromley, J.~M. Arrazola, M.~Schuld, N.~Quesada, and
  S.~Lloyd, ``Continuous-variable quantum neural networks,'' \emph{Physical
  Review Research}, vol.~1, no.~3, p. 033063, 2019.

\bibitem{huggins2019towards}
W.~Huggins, P.~Patil, B.~Mitchell, K.~B. Whaley, and E.~M. Stoudenmire,
  ``Towards quantum machine learning with tensor networks,'' \emph{Quantum
  Science and technology}, vol.~4, no.~2, p. 024001, 2019.

\bibitem{tacchino2020variational}
F.~Tacchino, P.~K. Barkoutsos, C.~Macchiavello, D.~Gerace, I.~Tavernelli, and
  D.~Bajoni, ``Variational learning for quantum artificial neural networks,''
  in \emph{2020 IEEE International Conference on Quantum Computing and
  Engineering (QCE)}.\hskip 1em plus 0.5em minus 0.4em\relax IEEE, 2020, pp.
  130--136.

\bibitem{nam2018automated}
Y.~Nam, N.~J. Ross, Y.~Su, A.~M. Childs, and D.~Maslov, ``Automated
  optimization of large quantum circuits with continuous parameters,''
  \emph{npj Quantum Information}, vol.~4, no.~1, pp. 1--12, 2018.

\bibitem{majumdar2021optimizing}
R.~Majumdar, D.~Madan, D.~Bhoumik, D.~Vinayagamurthy, S.~Raghunathan, and
  S.~Sur-Kolay, ``Optimizing ansatz design in qaoa for max-cut,'' \emph{arXiv
  preprint arXiv:2106.02812}, 2021.

\bibitem{verdon2019learning}
G.~Verdon, M.~Broughton, J.~R. McClean, K.~J. Sung, R.~Babbush, Z.~Jiang,
  H.~Neven, and M.~Mohseni, ``Learning to learn with quantum neural networks
  via classical neural networks,'' \emph{arXiv preprint arXiv:1907.05415},
  2019.

\bibitem{mitarai2018quantum}
K.~Mitarai, M.~Negoro, M.~Kitagawa, and K.~Fujii, ``Quantum circuit learning,''
  \emph{Physical Review A}, vol.~98, no.~3, p. 032309, 2018.

\bibitem{ostaszewski2021structure}
M.~Ostaszewski, E.~Grant, and M.~Benedetti, ``Structure optimization for
  parameterized quantum circuits,'' \emph{Quantum}, vol.~5, p. 391, 2021.

\bibitem{lloyd2018quantum}
S.~Lloyd, ``Quantum approximate optimization is computationally universal,''
  \emph{arXiv preprint arXiv:1812.11075}, 2018.

\bibitem{bosertraining}
B.~E. Boser, I.~M. Guyon, and V.~N. Vapnik, ``A training algorithm for optimal
  margin classifiers,'' in \emph{Proceedings of the 5th Annual ACM Workshop on
  Computational Learning Theory}, pp. 144--152.

\bibitem{chen2015xgboost}
T.~Chen, T.~He, M.~Benesty, V.~Khotilovich, Y.~Tang, H.~Cho \emph{et~al.},
  ``Xgboost: extreme gradient boosting,'' \emph{R package version 0.4-2},
  vol.~1, no.~4, 2015.

\bibitem{van2008visualizing}
L.~Van~der Maaten and G.~Hinton, ``Visualizing data using t-sne.''
  \emph{Journal of machine learning research}, vol.~9, no.~11, 2008.

\bibitem{elsken2019neural}
T.~Elsken, J.~H. Metzen, F.~Hutter \emph{et~al.}, ``Neural architecture search:
  A survey.'' \emph{J. Mach. Learn. Res.}, vol.~20, no.~55, pp. 1--21, 2019.

\bibitem{ren2020comprehensive}
P.~Ren, Y.~Xiao, X.~Chang, P.-Y. Huang, Z.~Li, X.~Chen, and X.~Wang, ``A
  comprehensive survey of neural architecture search: Challenges and
  solutions,'' \emph{arXiv preprint arXiv:2006.02903}, 2020.

\bibitem{wine}
U.~M.~L. Repository, ``Breast cancer and wine datasets,'' \emph{University of
  California, School of Information and Computer Science}.

\end{thebibliography}
\bibliographystyle{IEEEtran}

\appendices

\section{Experimental Environment and Setting}\label{apd:exp}
\begin{figure}[t]
    \centering
    \includegraphics[width = 0.48\textwidth]{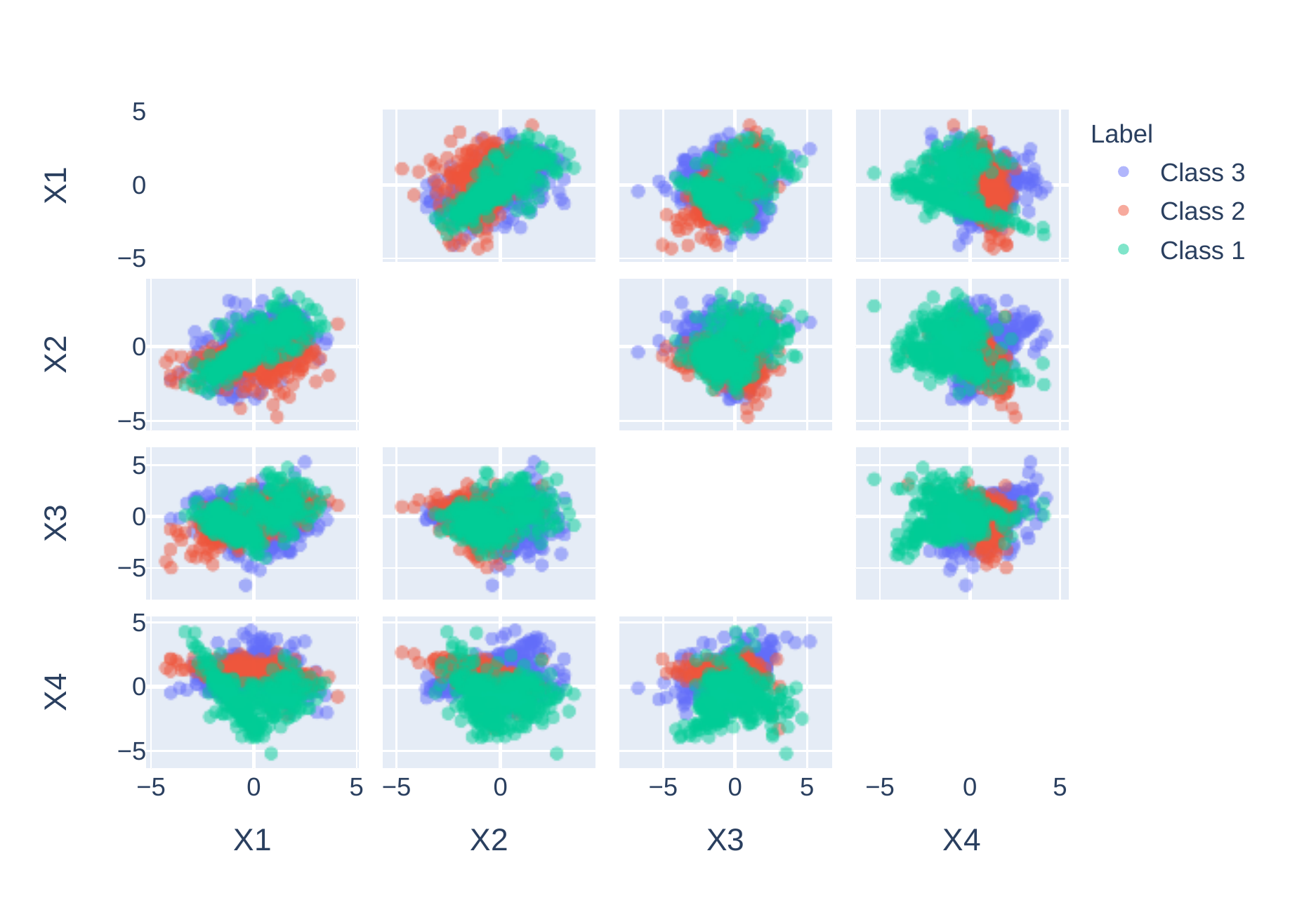}
    \includegraphics[width = 0.48\textwidth]{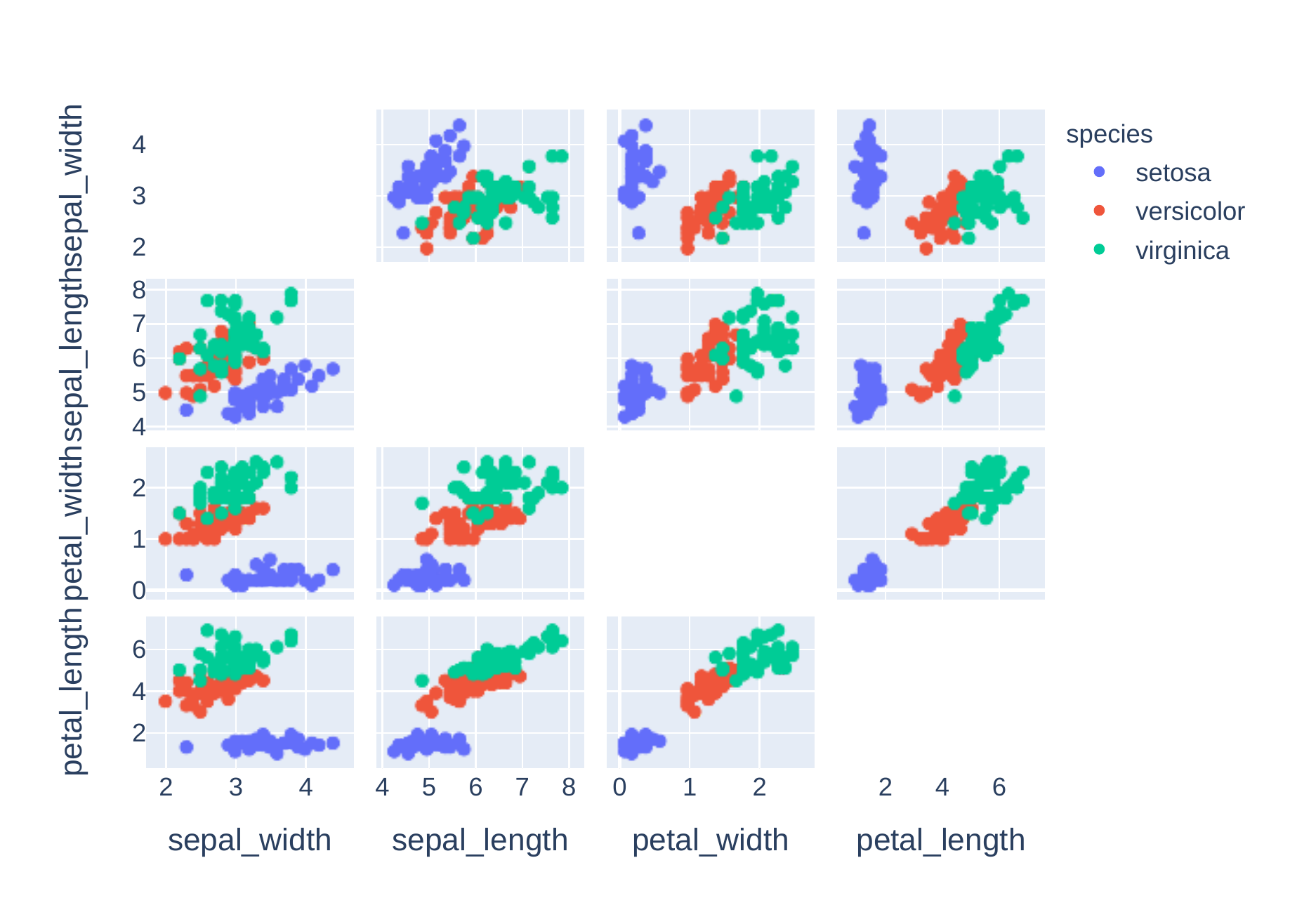}
    \caption{Illustrations of datasets used in experiments on stand-alone quantum embeddings: (Top) Synthesized dataset, (Bottom) Iris dataset.}
    \label{fig:low-dim-data}
\end{figure}
\begin{table}[t]
    \centering
    \begin{tabular}{c|c|c|c}
        \toprule
        \textbf{Dataset} & \textbf{\#Samples} & \textbf{\#Input \textbf{Features}} & \textbf{\#Classes}\\
        \midrule
        Wine\cite{wine} & $178$ & $13$ & $3$\\
        Breast Cancer\cite{wine} & $569$ & $30$ & $2$\\
        \bottomrule
    \end{tabular}
    \caption{Description of datasets used in experiments of hybrid classical-quantum neural architecture.}
    \label{tab:high-data}
\end{table}
Our experiments are conducted using Pennylane and Qiskit quantum simulators in Python 3.6 and Pytorch 1.8. The implementation of search strategies is via Optuna 2.7.0 package. For low dimensional datasets, we use Adam optimizer with initial learning rate of $0.5$ and $\beta = (0.9,0.999)$. The decay rate of the learning rate is set at $0.97$ with a decay period of every $2$ epoch. We train each architecture candidate for $100$ and $50$ epochs on synthesis and Iris datasets, respectively. Final validation accuracy of synthesis datasets based on five independent runs, while ten independent runs are used in the Iris dataset. For the hybrid neural architecture, we use SGD optimizer with learning rate of $0.5$ and momentum of $0.9$. The early stopping is used to yield final model, which monitors the improvement of validation loss over $10$ epochs. Regarding the parameter setting for the TPE sampler, we initialize the search phase with $20$ random trials, followed by $300$ trials. The number of samples to evaluate the expected improvement each trial is $1000$ samples. Moreover, the same number of trials is used for the random search sampler.

\section{Analysis of Preliminary Experiments}\label{apd:prelim}
\begin{table}[h]
    \centering
    \begin{tabular}{c|c|c}
    \toprule
    Layout 1 & Layout 2 & Layout 3 \\
    \midrule
    0.1965& 0.1341 & 0.1846\\
    0.1793& 0.1846 & 0.2011\\
    0.1513& 0.1184 & 0.2035\\
    0.1984& 0.1004 & 0.2151\\
    0.1765& 0.1048 & 0.2192\\
        
    \bottomrule     
    \end{tabular}
    \caption{The validation loss of preliminary quantum embedding layouts on Iris dataset. The same hyper-parameter setting is used across all experiments.}
    \label{tab:prelim-res}
\end{table}
In the experiment on the Iris dataset, we hand-crafted layouts of entanglements for observations. We start with sampling a quantum embedding with an entanglement level of $k=3$ to identify the first layout. Then we permute the order of CNOT gates to get the second layout. Finally, the last layout is constructed by adding a random CNOT gate to the second layout. The validation loss of each layout is reported in Table~\ref{tab:prelim-res}, which involves five independent runs. Let denote the actual means of the validation score as $\mu_1$, $\mu_2$, and $\mu_3$ for corresponding layouts. The $p$-value for the $t$-test of the null hypothesis $H_0: \mu_1 = \mu_2$ is $0.017$. Thus, we have strong statistical evidence to reject the null hypothesis, meaning the two means are not equal. Similarly, the $p$-value of the $t$-test over null hypothesis $H_0: \mu_2 = \mu_3$ is $0.001$, indicating that there are differences in the mean validation loss from different quantum embedding architectures.

\section{Description of Data scenarios}\label{apd:prelim}
\color{black}
We depict the low dimensional datasets used in experiments on stand-alone quantum embeddings in Figure~\ref{fig:low-dim-data}. The usage of small value for hypercube enhances the difficulty for simulated dataset, which leads to variations in performance when using different $k$. Table~\ref{tab:high-data} depicts the details of data used in hybrid classical-quantum neural architecture.
\color{black}



\end{document}